\documentclass[acmsmall,screen]{acmart}
\setcopyright{none}
\usepackage{amsmath}
\usepackage{amsfonts}
\usepackage{algorithm}
\usepackage{algorithmicx}
\usepackage{algpseudocodex}
\usepackage{multirow}
\usepackage{xspace}
\usepackage{array}
\usepackage{booktabs}
\usepackage{diagbox}
\usepackage{graphicx}
\usepackage{subfig}
\usepackage{textcomp}
\usepackage{url}
\usepackage{verbatim}
\usepackage{xcolor}
\usepackage{enumitem}
\usepackage{pifont}
\newcommand{\sysname}{{\sf LowDiff}\xspace}
\newcommand{\sysnameplus}{{\sf LowDiff+}\xspace}
\newcommand{\sysnamepluss}{{\sf LowDiff+(S)}\xspace}
\newcommand{\sysnameplusp}{{\sf LowDiff+(P)}\xspace}

\definecolor{LightBlue}{HTML}{4D88FF}

\makeatletter
\newcommand{\algsetline}[1]{\setcounter{ALG@line}{#1}}
\makeatother

\begin{document}

\title{Optimizing Frequent Checkpointing via Low-Cost Differential for Distributed Training Systems}

\titlenote{Extension of Conference Paper. This paper extends our conference paper~\cite{yao2025lowdiff} by more than 30\% new material: $(i)$ we add a parallel analysis showing that gradient reuse can run in parallel with training and be hidden within a training iteration; $(ii)$ we extend gradient reuse to training systems without gradient compression, where we identify two issues (snapshotting uncompressed gradients contends with training for bandwidth; persisting such large differentials saturates storage bandwidth) and two corresponding insights (gradients can be reused layer-by-layer once synchronized; differential checkpoints can be incrementally merged into a CPU-resident replica), and design and implement \sysnameplus, which employs layer-wise-reuse snapshotting and incremental-merging persistence to overlap checkpointing with training; and $(iii)$ we add a comprehensive evaluation on billion-parameter-scale models with fine-grained breakdowns of checkpointing and recovery, demonstrating the effectiveness of \sysname and \sysnameplus.}

\author{Chenxuan Yao}
\orcid{0000-0002-1143-052X}
\affiliation{\institution{Huazhong University of Science and Technology}\city{Wuhan}\country{China}}
\email{deadfffool@hust.edu.cn}

\author{Yuchong Hu}
\orcid{0000-0003-1265-7141}
\affiliation{\institution{Huazhong University of Science and Technology}\city{Wuhan}\country{China}}
\email{yuchonghu@hust.edu.cn}

\author{Feifan Liu}
\orcid{0009-0007-9790-7651}
\affiliation{\institution{Huazhong University of Science and Technology}\city{Wuhan}\country{China}}
\email{ffliu@hust.edu.cn}

\author{Zhengyu Liu}
\orcid{0009-0003-2432-0092}
\affiliation{\institution{Huazhong University of Science and Technology}\city{Wuhan}\country{China}}
\email{lawzyliu@hust.edu.cn}

\author{Lin Wang}
\orcid{0000-0001-8528-1232}
\affiliation{\institution{Huazhong University of Science and Technology}\city{Wuhan}\country{China}}
\email{wanglin2021@hust.edu.cn}

\author{Mingqi Li}
\orcid{0009-0001-7171-6925}
\affiliation{\institution{Huazhong University of Science and Technology}\city{Wuhan}\country{China}}
\email{mingqili2024@hust.edu.cn}

\author{Dan Feng}
\orcid{0000-0002-4674-6006}
\affiliation{\institution{Huazhong University of Science and Technology}\city{Wuhan}\country{China}}
\email{dfeng@hust.edu.cn}

\begin{abstract}
Distributed training of large deep-learning models often leads to failures, so checkpointing is commonly employed for recovery. State-of-the-art studies focus on frequent checkpointing for fast recovery from failures. However, frequent checkpointing generates numerous checkpoints, incurring substantial costs and thus degrading training performance. Recently, differential checkpointing has been proposed to reduce costs, but it is limited to recommendation systems, so its application to general distributed training systems remains unexplored.

In this paper, we find that gradients generated during distributed training can be reused to construct differential checkpoints, while the former's size is smaller than the latter's, motivating us to reuse gradients for low-cost differential checkpointing. Based on this main idea, we propose \sysname, a frequent checkpointing framework for compression-enabled training systems that reuses compressed gradients as differential checkpoints, eliminating redundant differential computation and reducing checkpoint transmission cost. Furthermore, we extend gradient reuse to scenarios without gradient compression and propose \sysnameplus, which employs layer-wise-reuse snapshotting and incremental-merging persistence to overlap checkpointing with training execution. Experiments on diverse workloads, including billion-parameter-scale models, demonstrate that \sysname and \sysnameplus significantly reduce checkpointing overhead and enable checkpointing at frequencies as high as once per iteration, reducing training time by up to 89.2\% and 81.2\%, respectively.

\end{abstract}




\maketitle

\section{Introduction}

Deep learning has achieved significant advances in computer vision \cite{simonyan2014very,he2016deep}, natural language processing \cite{devlin2018bert,brown2020language} and so on. To efficiently train deep learning models, \textit{distributed training systems} \cite{sergeev2018horovod,rasley2020deepspeed,li2020pytorch} are widely deployed across workers to accelerate the training process. Nevertheless, many workers and long training time required by distributed training may lead to frequent failures \cite{zhang2020empirical}. For instance, OPT-175B, equipped with 992 NVIDIA A100 GPUs, experienced 112 failures during a 90-day training period, averaging nearly two failures per day \cite{OPT}.

To handle failures, \emph{checkpointing}, which regularly writes the model state to storage for failure recovery, can confine the potential loss of the training progress to the interval between consecutive checkpoints, only requiring the recovery of the remaining progress. Traditional checkpointing saves the model states (including the model parameters and optimizer parameters) at epoch boundaries, but an epoch often runs for hours, so the epoch-level checkpointing often results in hours of GPU computation wasted on average for every failure recovery \cite{mohan2021checkfreq}.

Consequently, state-of-the-art checkpointing studies on fast recovery (e.g., Checkfreq \cite{mohan2021checkfreq} and Gemini \cite{wang2023gemini}) aim to increase the frequency of checkpointing from epochs to iterations (we call \textit{frequent checkpointing}). Nevertheless, the frequent checkpointing generates a considerable number of checkpoints and incurs substantial communication costs, thus stalling the training process. Recent studies on reducing such costs (e.g., Check-N-Run \cite{eisenman2022check}) introduce \emph{differential checkpointing} (\textit{DC} for short), which only tracks and checkpoints the modified part of the model to reduce the checkpoint size, into recommendation systems by leveraging its unique sparse features.

However, when deploying DC with high frequency in general distributed training systems, we observe two challenges arising: $i)$ without the help of sparse features, frequent DC incurs significant computation cost required for compressing the original differentials into smaller ones (i.e., \textit{compressed differentials}), thus hindering the training process, and $ii)$ frequent DC causes significant training stalls, as frequent checkpoint writes incur substantial transmissions which cannot be overlapped with training, thus degrading the training performance. 

We observe that gradients generated during distributed training naturally provide opportunities for reducing DC overhead. Specifically, we have two findings: $i)$ gradients can be used to construct differential checkpoints for general deep-learning models, and $ii)$ gradients are much smaller than differential checkpoints. When gradient compression is applied, these findings further indicate that compressed gradients can represent compressed differentials with a substantially smaller size.

Based on these findings, we propose our main idea of gradient reuse. In our conference work~\cite{yao2025lowdiff}, we proposed \sysname, which utilizes this idea in training systems that employ gradient compression \cite{chen2023bbtopk,fei2021efficient,ming2024adtopk,song2023optimus}, a technique widely used for communication efficiency. \sysname is a low-cost differential checkpointing framework that reuses compressed gradients as differential checkpoints, thereby eliminating differential-compression overhead and reducing checkpoint transmission cost. 

We further extend the main idea of gradient reuse to scenarios without gradient compression. We first identify the issues of snapshotting and persistence when dealing with uncompressed gradients. To address these, we offer two key insights: $i)$ gradients can be reused layer-by-layer, and $ii)$ differential and full checkpoints can be merged together. Based on the two insights, we propose \sysnameplus, which employs a layer-wise-reuse snapshotting strategy, along with an incremental-merging persistence strategy, thus enabling high-frequency checkpointing without gradient compression.

We evaluate \sysname and \sysnameplus on diverse workloads with billion-parameter models. The results show that both systems support checkpointing as frequently as once per iteration with minimal runtime overhead: \sysname and \sysnameplus reduce the training time of large models by up to 89.2\% and 81.2\% compared with state-of-the-art checkpointing systems, respectively. 
\section{Background and Related Work}

\subsection{Basics of Distributed Training Systems}
\label{subsec:dnn}
Training a deep neural network (DNN) proceeds iteratively over a dataset many times (i.e., \textit{epochs}), each of which contains multiple \textit{iterations}, with an \textit{optimizer} adjusting the model \textit{parameters} based on the \textit{gradients} (i.e., derivatives of the loss function w.r.t. model parameters).

To accelerate the training process of large models and datasets \cite{zhao2023pytorch,li2020pytorch,sergeev2018horovod,rasley2020deepspeed},  \textit{distributed training systems} are commonly adopted to distribute computations and data across multiple GPUs. Specifically, distributed training has four steps for the $t$-th iteration:
\begin{itemize}[leftmargin=10pt]
\item\emph{Forward pass.} The model parameters $x_t$ are applied to input data $X$ for prediction $Y$ as:
\begin{equation}
\label{eqn:forward}
Y = Forward(x_t,X). 
\end{equation}
\item\emph{Backward pass.} A loss function is used to calculate the gradient $G_{i,t}$ on $i$-th worker as:
\begin{equation}
\label{eqn:backward}
G_{i,t} = Backward(x_t,Y).
\end{equation}
\item\emph{Synchronization.} The gradients of workers are synchronized through communication as:
\begin{equation}
\label{eqn:sync}
G_{t} = Sync(G_{i,t}).
\end{equation}
\item\emph{Model update.} The optimizer (e.g., Adam \cite{kingma2014adam}) uses the synchronized gradient $G_t$ to update the \textit{model state} (denoted by $M_t$), which contains the model parameters $x_t$ and the optimizer parameters $o_t$ (i.e., $M_{t} = (x_t ,o_t)$), satisfying that 
\begin{equation}
\label{eqn:update}
M_{t+1} \gets M_{t} + Adam(G_t). 
\end{equation}
\end{itemize}
Here, we specially introduce the widely used optimizer, Adam. The Adam optimizer maintains the first-order and second-order moments (i.e., the mean and variance of the gradients), and each has the same size as the model parameters, denoted by $\Psi$. Therefore, Adam requires an additional $2\Psi$ parameter storage beyond the model itself, which will support one of our findings in \S\ref{subsec:insight}.

Distributed training systems typically operate under two contrasting regimes: one employing \textit{gradient compression} (see \S\ref{subsec:compression}) under communication constraints \cite{vogels2019powersgd,song2023optimus}, and the other foregoing compression via high-bandwidth interconnects such as NVLink \cite{narayanan2021efficient,megascale}. In this paper, we propose our system designed toward both scenarios (see \S\ref{sec:design} and \S\ref{sec:enhance}).

\subsection{Checkpointing Techniques}
\label{subsec:ckpt_back}
When training large models, failures frequently occur in distributed training systems due to extensive computational resources and long training times. To handle frequent training failures, \textit{checkpointing}, a widely used technique in common training frameworks such as PyTorch \cite{paszke2019pytorch}, has been extensively studied  \cite{eisenman2022check,mohan2021checkfreq,wang2023gemini,gupta2024just,wang2024fastpersist,chen2023cost,maurya2024datastates}. It regularly copies the model state to local storage or remote storage during training. During recovery from the failure, all GPUs load the latest checkpoint and rerun the remaining process until the point of failure. To evaluate checkpointing performance, recent studies (e.g., \cite{gupta2024just}) define the \emph{wasted time} metric as the sum of the recovery time from the latest checkpoint and the steady-state overhead (i.e., the GPU time for checkpointing when no failures have happened). 

\noindent\textbf{Frequent checkpointing.} Traditional checkpointing is performed at the granularity of epochs, but an epoch may run for hours, thus incurring substantial wasted time for every failure \cite{mohan2021checkfreq}. Thus, state-of-the-art studies \cite{mohan2021checkfreq,wang2023gemini} that aim for fast recovery increase the checkpointing frequency from epochs to iterations, which we call \textit{frequent checkpointing} in this paper. 

Checkfreq \cite{mohan2021checkfreq} pipelines GPU-to-CPU \textit{snapshot} and asynchronous \textit{persist} operations with computation, creating a checkpoint every 14–19 iterations. Gemini \cite{wang2023gemini} exploits host CPU memory with checkpointing traffic scheduling, increasing the checkpoint frequency by up to $8\times$ over Checkfreq, even per iteration. PCcheck \cite{strati2025pccheck} checkpoints to persistent memory with concurrent checkpoints and multi-threaded pipelining, achieving a checkpoint every 10 iterations. Just-in-time checkpointing \cite{gupta2024just} initiates a checkpoint only when a failure is detected, assuming redundancy across workers.

Despite the fast recovery enabled by frequent checkpointing, the creation and transmission costs may not be fully hidden by asynchronous checkpointing when checkpoints are generated every few iterations or even every iteration, thus degrading training performance. In this paper, our main goal is to lower the costs of frequent checkpointing for high-performance training.

\noindent\textbf{Differential checkpointing.}
A recent study, Check-N-Run \cite{eisenman2022check}, on reducing the checkpointing costs introduces \emph{differential checkpointing} (\textit{DC} for short), which records the part of the model that is modified only during the last checkpoint interval so as to reduce the checkpoint size. 

Specifically, DC saves its model state as \textit{full checkpoint} ($C^F$) regularly. DC also saves its \textit{differential checkpoint} of the $t$-th iteration (denoted by $C^D_{t}$) at an iteration-level high frequency as 
\begin{equation}
\label{eqn:diff}
C^D_{t} = M_{t+1}-M_{t}.
\end{equation}
Then the model state $M_{t}$ at the $t$-th iteration can be reconstructed by merging the full and differential checkpoints as
\begin{equation}
\label{eqn:dc}
M_{t} = C^{F} + C^D_{1} + C^D_{2} + \dots + C^D_{t-1}.
\end{equation}
Namely, $C^D_t$ is produced at the $t$-th iteration and carries the state transition from $M_t$ to $M_{t+1}$, so the sum in Equation~(\ref{eqn:dc}) runs up to $C^D_{t-1}$, the last differential needed to reach $M_t$.

However, Check-N-Run deploys DC only in recommendation systems, where the sparse embedding table ensures that only a \textit{small part} of the model parameters is updated per iteration. General DNN models instead have to be updated \textit{entirely} per iteration, so deploying DC has to highly compress the differentials, incurring substantial cost that grows with frequency (see \S\ref{subsec:observations}).

In this paper, we focus on how to enable low-cost DC for general DNN models, so as to extend the DC from recommendation systems to general distributed training systems.

\subsection{Gradient Compression Techniques}
\label{subsec:compression}
\emph{Gradient compression} techniques have been widely used in high-performance distributed training systems \cite{aji2017sparse,fang2019redsync,m2021efficient,shi2019distributed,stich2018sparsified,dettmers20158,alistarh2017qsgd,song2023optimus} to significantly reduce the communication overhead of the gradient synchronization in backward pass (see \S\ref{subsec:dnn} and Equation~(\ref{eqn:sync})). Two representative compression approaches are \emph{Sparsification} \cite{aji2017sparse,stich2018sparsified}, which selects only a subset of the elements of the gradient, and \emph{Quantization} \cite{dettmers20158,alistarh2017qsgd}, which reduces the number of bits of each element of the gradient.

In this paper, we show that training gradients can construct differential checkpoints, motivating gradient reuse, with gradient compression providing an efficient realization (see \S\ref{subsec:motivation}).

\section{Observations and Motivation}
\label{sec:observation}

\begin{figure}[t]
    \centering
    \subfloat[Computation Frequency\label{fig:compute_overhead}]{
        \includegraphics[width=0.26\linewidth]{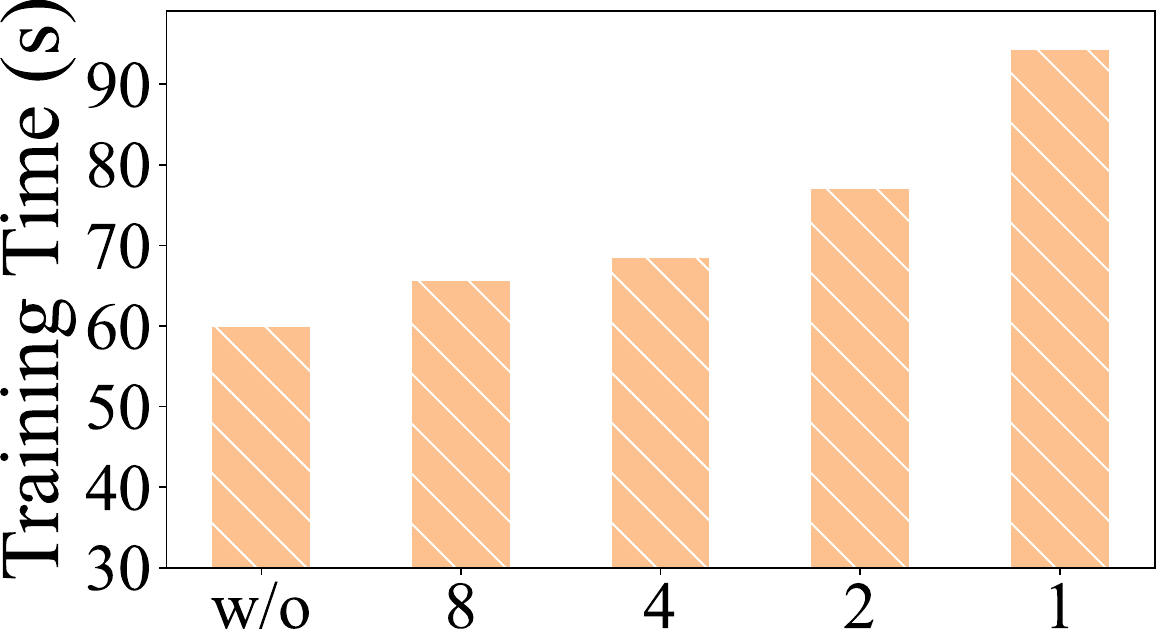}
    }
    \hfill
    \subfloat[Transmission Frequency\label{fig:transmit_overhead}]{
        \includegraphics[width=0.26\linewidth]{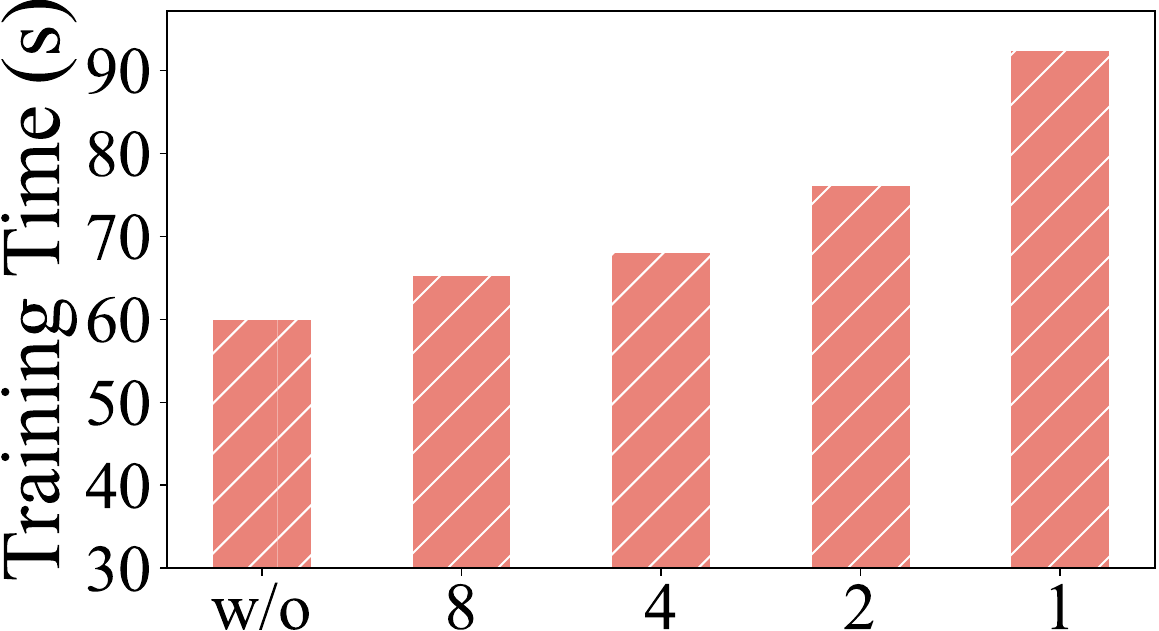}
    }
    \hfill
    \subfloat[Equivalence to DC\label{fig:finding1}]{
        \includegraphics[width=0.22\linewidth]{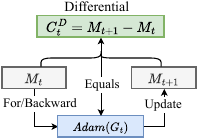}
    }
    \hfill
    \subfloat[Size reduction\label{fig:finding2}]{
        \includegraphics[width=0.16\linewidth]{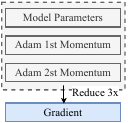}
    }
    \vspace{-10pt}
    \caption{Challenges of frequent DC (a--b) and findings from gradients and DCs (c--d).}
    \Description{Challenges of frequent DC and findings from gradients and DCs.}
    \label{fig:comparison}
    \vspace{-10pt}
\end{figure}

In this section, we present two challenges of deploying frequent differential checkpointing (DC) in distributed training systems (\S\ref{subsec:observations}) and discuss two corresponding findings from the relationship between gradients and DCs (\S\ref{subsec:insight}). Finally, we demonstrate how these observations motivate our main idea to address the challenges (\S\ref{subsec:motivation}).

\subsection{Challenges of DC in General Training}
\label{subsec:observations}

As stated in \S\ref{subsec:ckpt_back}, DC may incur high costs when deployed in general distributed training systems without the help of the sparsity feature in recommendation systems. Therefore, we conduct measurements to show the challenges of DC in general distributed training systems, especially under high frequency. We conduct our measurement on eight NVIDIA A100 GPUs with a 25Gbps network \cite{wang2023gemini}. We evaluate a typical large-model training task GPT2-L \cite{radford2019language}, with a common differential checkpointing scheme (see Equation (\ref{eqn:dc})). Detailed setups are shown in \S\ref{subsec:setup}.

\noindent\textbf{Challenge 1: Frequent DC incurs significant computation cost for highly compressing differentials.} In the absence of the sparsity feature's support, DC has to apply the highly compressed scheme (compression ratio $\rho$ = 0.01 \cite{eisenman2022check}) to the differentials (called \textit{compressed differentials}), thus degrading the training performance. To measure the impact of compression computation cost on training, Figure \ref{fig:compute_overhead} shows the training time of cases without compression and with different compression frequencies (e.g., 8 iteration compression frequency means the compression is performed every 8 iterations). Compared to the no compression case, the compression case slows down training by 13\%–57\%, while higher frequency incurs much slower training. 

\noindent\textbf{Challenge 2: Frequent DC causes significant transmission cost due to frequent checkpoint writes.} Frequent DC needs to write the differential of model state to storage frequently, which blocks the training process frequently and incurs substantial transmission overhead, since the model update has to wait for the snapshotting of checkpoint \cite{mohan2021checkfreq,wang2024fastpersist}. To measure the impact of transmission cost on training in terms of frequency, Figure \ref{fig:transmit_overhead} shows the training time without differential transmission and with different differential transmission frequencies. We observe that the case with differential transmission slows down the training process by 12\%–54\% compared to that without differential transmission, while higher frequency incurs slower training.

\subsection{Findings from Gradients and Differentials}
\label{subsec:insight}

To address the above challenges, we resort to the gradients in distributed system, and obtain the following two findings.

\noindent\textbf{Finding 1: Gradients can directly construct the differentials for general DNN models.} According to Equation (\ref{eqn:update}), the model state updated by gradient $G_t$ can be represented as $Adam(G_t) = M_{t+1}-M_{t}$, which corresponds to the DC defined in Equation (\ref{eqn:diff}) as $C_t^D = M_{t+1}-M_{t}$. Thus, we have:
\begin{equation}
\label{equ:ada}
C_t^D = Adam(G_t). 
\end{equation}
 As shown in Figure \ref{fig:finding1},  a gradient can therefore construct the corresponding DC through the optimizer (e.g., Adam). Furthermore, when the same compression method is applied, the corresponding compressed DC can be constructed from the compressed gradient.

\noindent\textbf{Finding 2: Gradients are one-third the size of differential checkpoints.} 
As mentioned in \S\ref{subsec:dnn}, if the model parameters are of size $\Psi$, the Adam optimizer parameters occupy $2\Psi$, making a full checkpoint $3\Psi$ in total. Next, Equation (\ref{eqn:dc}) implies that a DC has the same size as the full one, i.e., $3\Psi$, which includes both model and Adam optimizer parameters. By contrast, the size of a gradient is equal to that of the model parameters, i.e., $\Psi$. Figure~\ref{fig:finding2} illustrates the size relationship between gradients and DCs. Furthermore, applying the same compression method results in a compressed gradient three times smaller than a compressed DC.

\subsection{Main Idea}
\label{subsec:motivation}

Based on Findings 1 and 2, our main idea is to \textit{reuse} gradients already generated by the training process to construct differential checkpoints. This gradient reuse principle avoids redundant differential computation and reduces checkpoint data movement. Therefore, we propose \sysname, a low-cost differential checkpointing framework that reuses compressed gradients as differential checkpoints. We further
extend gradient reuse to scenarios without gradient compression and propose \sysnameplus (See \S\ref{sec:enhance}).

\begin{figure*}[t]
    \centering
    \includegraphics[width=0.9\textwidth]{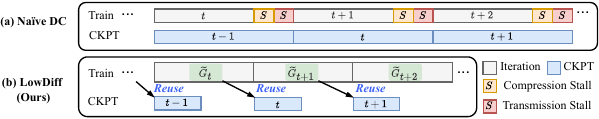}
    \vspace{-10pt}
    \caption{Motivating example of \sysname.}
    \Description{Motivating example of \sysname.}
    \label{fig:motivation}  
    \vspace{-10pt}
\end{figure*}

As illustrated in Figure~\ref{fig:motivation}(b), \sysname reuses
$\widetilde{G}_{t}$, $\widetilde{G}_{t+1}$, and $\widetilde{G}_{t+2}$ as DCs, eliminating the compression stalls (Challenge~1) and reducing the transmission stalls (Challenge~2) shown in Figure~\ref{fig:motivation}(a). The feasibility of executing \sysname concurrently with training requires further analysis. We next examine its data dependencies and overlap opportunities in \S\ref{sec:ana}.

\section{Analysis}
\label{sec:ana}

By incorporating the main idea of gradient reuse (\S\ref{sec:observation}), \sysname aims to run checkpointing concurrently with model training, thereby minimizing potential stalls. We confirm the feasibility from two aspects: whether data dependencies block the concurrency (\S\ref{subsec:para}), and whether the checkpointing cost can be hidden within a training iteration (\S\ref{subsec:overlap}).

\subsection{Checkpointing–Training Parallelism}
\label{subsec:para}

\begin{figure}[t]
    \centering
    \includegraphics[width=0.92\linewidth]{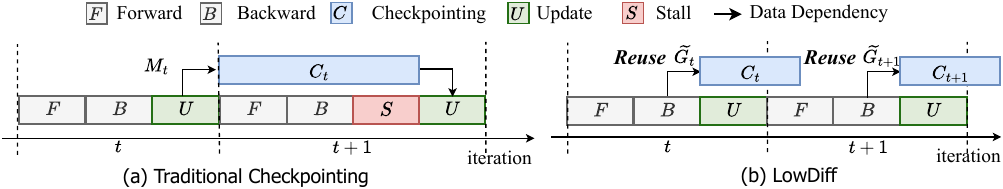}
    \vspace{-10pt}
    \caption{\sysname can run in parallel with forward pass ($F$), backward pass ($B$), and model update ($U$).}
    \Description{\sysname can run in parallel with forward pass ($F$), backward pass ($B$), and model update ($U$).}
    \label{fig:ob3}  
    \vspace{-10pt}
\end{figure}

\noindent\textbf{Necessity of Parallelism Analysis}. Traditional checkpointing allows an overlap with the forward and backward passes, which do not modify the model state, but it cannot run concurrently with the model update~\cite{mohan2021checkfreq, wang2024fastpersist}. Checkpointing must read the updated model state, creating a RAW (Read-after-Write~\cite{hennessy2011computer}) dependency with the current model update and a WAR (Write-after-Read~\cite{hennessy2011computer}) dependency with the next update. As illustrated in Figure~\ref{fig:ob3}(a), these dependencies introduce synchronization stalls and constrain checkpointing--training parallelism. We therefore analyze whether \sysname's gradient-reuse strategy removes these dependencies and enables checkpointing to execute concurrently with training.

\noindent\textbf{\sysname can eliminate the WAR dependency.} We next analyze the WAR dependency with the next iteration's model update, backward, and forward pass.
\begin{itemize}[leftmargin=10pt]
    \item \textbf{Model update.} As described in \S\ref{subsec:dnn}, the model update (Equation~(\ref{eqn:update})) only requires read-only access to the gradient. Because \sysname checkpoints only the compressed gradient, the original Write-after-Read (WAR) dependency is eliminated because the checkpointing process no longer needs to read the model state ($M_t$). Instead, it just reads the compressed gradient ($\widetilde{G}_t$), which is available in the backward pass and will not be changed after. Since both the checkpointing and model update are read-only consumers of the gradient, they can operate in parallel without conflict.
    \item \textbf{Backward pass.} \sysname's gradient reuse strategy introduces no additional WAR dependency on backward pass, since each gradient ${G}_t$ is finalized within its iteration and never modified thereafter. Specifically, once a gradient ${G}_t$ is generated in the current iteration ($t$), it is final and will not be modified by subsequent steps. The backward pass of the $(t+1)$-th iteration (Equation~(\ref{eqn:backward})) does not read or depend on ${G}_t$; it generates a completely new gradient $G_{t+1}$ from scratch and won't change ${G}_t$. Therefore, the DC's read of ${G}_t$ does not conflict with the backward pass's write of ${G}_{t+1}$ at the $(t+1)$-th iteration, allowing \sysname to run in parallel with the next iteration's backward pass. 
    \item \textbf{Forward pass.} The forward pass (Equation~(\ref{eqn:forward})) doesn't involve the gradient, so there is no WAR dependency between \sysname and the forward pass.
\end{itemize}

\noindent\textbf{\sysname can run in parallel with forward pass, backward pass, and model update.} To visualize the parallelism benefit from data dependency, Figure \ref{fig:ob3}(b) shows the execution timeline with \sysname. We see that the checkpointing ($C_t$) at the $t$-th iteration is no longer coupled with the model update ($U$) or the following iteration (no dependency arrow from checkpointing to training). Instead, the reusing operation on compressed gradient ($\widetilde{G}_t$) begins immediately after the backward pass ($B$), running in parallel with the model update and subsequent iteration's forward pass ($F$), backward pass ($B$), and model update ($U$). This parallelism property effectively transforms \sysname from a blocking operation to a parallel execution with training, thereby eliminating the synchronization stalls previously caused by the WAR dependency.

\begin{figure}[t]
    \centering
    \includegraphics[width=0.6\linewidth]{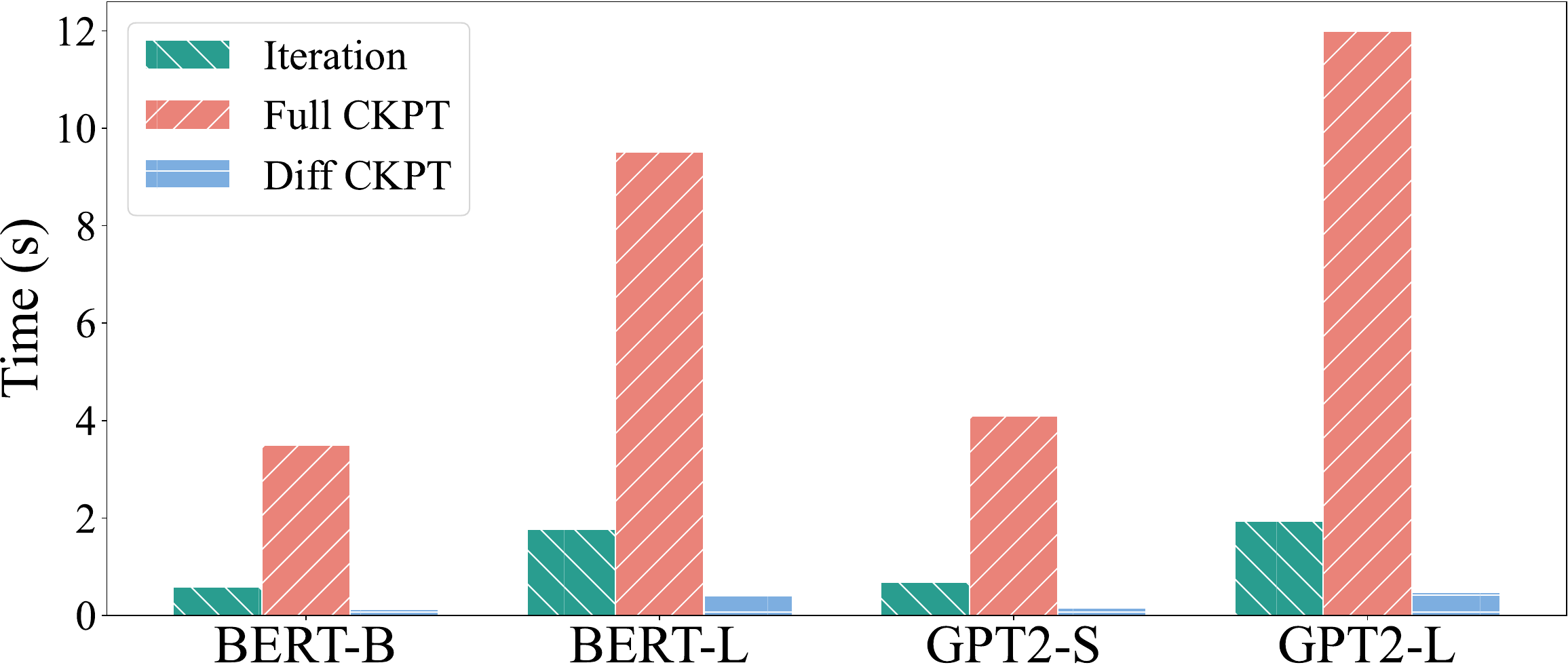}
    \vspace{-10pt}
    \caption{Time of iteration, full checkpointing, and differential checkpointing.}
    \Description{Time of iteration, full checkpointing, and differential checkpointing.}
    \label{fig:analysis}
    \vspace{-10pt}
\end{figure}

\subsection{Overlap Analysis}
\label{subsec:overlap}

\noindent\textbf{Necessity of Overlap Analysis}. Beyond its parallelism property, \sysname can perform even better if the checkpointing process is fast enough to be hidden by the training iteration, thus minimizing DC's cost. In other words, the time required to checkpoint a compressed gradient (denoted as $C$) should be shorter than a single training iteration (denoted as $F+B+U$), similar to the Microsoft study~\cite{mohan2021checkfreq}.

\noindent\textbf{The checkpointing of compressed gradients can be overlapped with training.} Following the measurement setup in \S\ref{subsec:observations}, we measure the iteration time, full checkpointing time, and differential checkpointing time on BERT-B, BERT-L, GPT2-S, and GPT2-L. As shown in Figure~\ref{fig:analysis}, the DC time is only 20.5\%, 22.1\%, 23.1\%, and 24.6\% of the corresponding training iteration time for these models. This indicates that in most cases, the checkpointing process can be fully overlapped with training iterations without introducing noticeable stalls. 

\noindent\textbf{Modern hardware makes the overlap practically feasible.} NVMe SSDs based on PCIe~4.0 offer write speeds around 5~GB/s (and up to 12~GB/s for PCIe~5.0, with RAID~0 configurations being even faster). Given that \sysname's compressed-gradient checkpoints are relatively small (several hundred MBs to a few GBs, as evaluated in \S\ref{sec:evaluation}), the theoretical I/O write time is typically under 1~second. This duration is significantly shorter than that of a full training iteration \cite{wang2023gemini}. 

Based on the above analysis, we confirm that the overlap of concurrent executions of \sysname's DC and training is practically feasible. 

\section{Design}
\label{sec:design}

\begin{figure*}[t]
    \centering
    \includegraphics[width=0.95\textwidth]{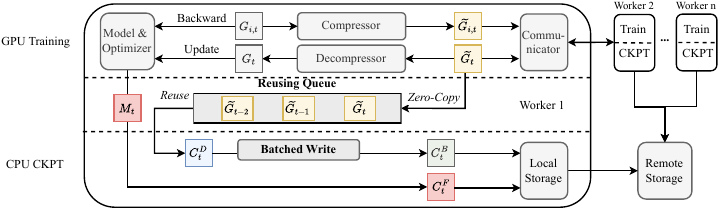}
    \vspace{-10pt}
    \caption{Architecture of \sysname.}
    \Description{Architecture of \sysname.}
    \label{fig:architecture}
    \vspace{-15pt}
\end{figure*}

Based on the gradient reuse main idea introduced in \S\ref{subsec:motivation}, we design \sysname with two components:
$(i)$ a gradient reuse module for constructing differential checkpoints without additional compression, and
$(ii)$ a batched writing module for efficient persistence under high checkpoint frequency.

Figure~\ref{fig:architecture} shows the architecture of \sysname. \sysname decouples checkpointing from training by introducing an independent \texttt{Checkpointing} process, which reuses synchronized compressed gradients from the original training pipeline as differential checkpoints. The \texttt{Training} and \texttt{Checkpointing} processes communicate through a \texttt{Reusing Queue} using zero-copy gradient transfer. The checkpointing process asynchronously consumes compressed gradients to construct differential checkpoints, which are further aggregated by the \texttt{Batched Writing} module before persistence.

\subsection{Reusing Compressed Gradients}
\label{subsec:LowDiff-p}

\noindent\textbf{Design requirements.} 
To reuse compressed gradients as differential checkpoints, \sysname needs to: $i)$ preserve the sequential order of gradients generated during training for correct differential replay, and $ii)$ efficiently transfer gradients between training and checkpointing processes with low communication overhead.

\noindent\textbf{Reusing queue structure.}
To meet requirement $i$, we design a queue-based data structure, referred to as the compressed gradient \emph{Reusing Queue} in Figure~\ref{fig:architecture}. Specifically, the first-in-first-out (FIFO) property of the queue naturally aligns with the iterative nature of training and checkpointing, ensuring that gradients are reused in the correct order. In addition, the implementation of the queue structure in PyTorch also helps meet requirement $ii$, which will be specified below.

\noindent\textbf{Reusing with zero-copy optimization.}
To meet requirement $ii$, we use the CUDA Inter-Process Communication (CUDA IPC) queue in PyTorch to implement a zero-copy mechanism, which allows GPU memory to be shared between training and checkpointing without extra copying. Specifically, the queue serves as a high-level data structure for reusing compressed gradients, but only transmits the memory handle. It relies on the CUDA IPC library to share the gradient's GPU memory handle across processes, avoiding additional I/O overhead from memory copying.

\begin{algorithm}[t]
    \caption{\sysname with an existing gradient-compressed training process}
    \label{alg:computation-free}
    \small
    \begin{algorithmic}[1]
    \Require{Model state $M_t$ at iteration $t$, Gradient $G_{i,t}$ at worker $i$, Reusing queue $Q$, Full checkpoint $C^F_t$, Differential checkpoints $\{C^D_j\}_{j=t}^{n-1}$, where $n-1$ is the last iteration at which a differential checkpoint is persisted (i.e., $M_n$ is the latest recoverable state), Compressor $Comp$, Decompressor $Comp^{-1}$.}
    \end{algorithmic}
    \vspace{2pt}
    \noindent\begin{minipage}[t]{0.49\linewidth}%
    \begin{algorithmic}[1]
    \Function{Training Process}{}
        \ForAll{$t$}
            \State $G_{i,t} \gets Backward(x_t)$ \Comment{Backward Pass}
            \State $\widetilde{G}_{i,t} \gets Comp(G_{i,t})$ \Comment{Compress}
            \State $\widetilde{G}_{t} \gets Sync(\widetilde{G}_{i,t})$ \Comment{Synchronize}
            \State $\textcolor{LightBlue}{\textbf{Q.put}}(\widetilde{G}_{t})$ \Comment{Zero-Copy}
            \State $G_{t} \gets Comp^{-1}(\widetilde{G}_{t})$ \Comment{Decompress}
            \State $M_{t+1} \gets M_{t} + Adam(o_t,G_t)$  \Comment{Update}
        \EndFor
    \EndFunction
    \end{algorithmic}
    \end{minipage}\hfill
    \begin{minipage}[t]{0.49\linewidth}%
    \begin{algorithmic}[1]
    \algsetline{8}
    \Function{Checkpointing Process}{}
        \While{$\widetilde{G}_{t} \gets \textcolor{LightBlue}{\textbf{Q.get}}$}  \Comment{Reusing}
            \State $C^D_t \gets Save(\widetilde{G}_{t})$  \Comment{Diff CKPT}
        \EndWhile
        \State $C^F_t \gets Save(M_{t})$  \Comment{Full CKPT Regularly}
    \EndFunction
    \Function{Recovery Process}{}
        \State $M_t \gets Load(C^F_t)$    \Comment{Load Full CKPT}
        \For{$j \gets t$, \dots, $n-1$}  \Comment{Pipelined Replay}
            \State $\widetilde{G}_j \gets Load(C^D_j)$  \Comment{Stage 1: Load}
            \State $G_j \gets Comp^{-1}(\widetilde{G}_j)$ \Comment{Stage 2: Decomp.}
            \State $M_{j+1} \gets M_{j} + Adam(G_j)$  \Comment{Stage 3: Merge}
        \EndFor
    \EndFunction
    \end{algorithmic}
    \end{minipage}
\end{algorithm}

\noindent\textbf{Design details.} 
Algorithm~\ref{alg:computation-free} shows the training, checkpointing, and recovery workflows of \sysname.

For the training process, after performing a backward pass and generating the gradient $G_{i,t}$ of worker $i$ at the $t$-th iteration (Lines 1-3), the training process compresses $G_{i,t}$ into $\widetilde{G}_{i,t}$ for communication (Line 4). Then, the compressed gradient is synchronized across all workers using a collective communication primitive (e.g., Allgather or Allreduce) (Line 5). Following synchronization, the training process puts the compressed gradient $\widetilde{G}_{t}$ into \texttt{Reusing Queue} $Q$ (Line 6), which will be transmitted to the checkpointing process with \textit{zero-copy}. Finally, the training process decompresses the gradient (Line 7), updates the model state (Line 8), and proceeds to the next iteration.

For the checkpointing process, it repeatedly retrieves the compressed gradients from \texttt{Reusing Queue} $Q$ to CPU memory (Line 10). Subsequently, the checkpointing process determines whether to merge the compressed gradient into batched differential checkpoints (detailed in \S\ref{subsec:LowDiff-b}). After that, it invokes $\texttt{torch.save}$ to persist the compressed differential checkpoints $C^D_t$ (Lines 11) to storage. The model state is also regularly checkpointed as a full checkpoint $C^F_t$ at the $t$-th iteration (Line 12). 

During recovery, the recovery process first loads the latest full checkpoint $C^F_t$ to restore the model state $M_t$ (Line 14), then replays the differential checkpoints generated after iteration $t$, i.e., $C^D_j$ for $j \in [t, n-1]$ (Lines 15-18): for each $C^D_j$, it loads the compressed gradient $\widetilde{G}_{j}$ (Line 16), decompresses it into $G_{j}$ (Line 17), and merges it into the model state via the optimizer and gradient (Line 18), until the training state is finally restored to $M_{n}$. These three stages are pipelined across consecutive differential checkpoints: while one checkpoint is being merged, the next is being decompressed, and the subsequent one is being asynchronously read from storage. The pipelining overlaps I/O with computation, leaving only the merge operations on the recovery critical path, thereby reducing the exposed recovery latency without changing the reconstruction semantics.

\noindent\textbf{Implementation.}
\sysname uses \texttt{torch.multiprocessing} with the \texttt{spawn} method for process isolation, bypassing Python's GIL. The \texttt{Reusing Queue} is built on \texttt{torch.multiprocessing.Queue()}, whose inter-process communication creates a CUDA IPC handle when a gradient tensor is enqueued and reopens it when dequeued, thereby achieving zero-copy GPU memory sharing while preserving the order of differential checkpoints.

\subsection{Batched Gradient Writing Optimization}
\label{subsec:LowDiff-b}

\noindent\textbf{Design requirements.} We find that current \sysname in \S\ref{subsec:LowDiff-p} still has two limitations: $i)$ frequent differential checkpoints incur significant write overhead due to the large number of writes, and $ii)$ avoiding the WAR dependency requires temporarily retaining the compressed gradient $\widetilde{G}_t$ for each DC operation, which causes the gradient buffer to remain occupied during storage writes and leads to extra GPU memory consumption, as analyzed in \S\ref{subsec:para}.

\noindent\textbf{Batched write.} To address limitation $i$, we can batch multiple writes of compressed gradients (reused as differential checkpoints) into a single write operation, with the help of the widely-used gradient accumulation technique \cite{hermans17a, accumulation} where the gradients of the same shape and size can be accumulated. 
Unlike DC, traditional full checkpointing cannot benefit from batching, since previous full checkpoints become obsolete once a new one is saved.

\noindent\textbf{Offloading batching to CPU.} To address limitation $ii$, we adopt a classic D2H offloading scheme \cite{rajbhandari2020zero} that offloads the batching operation to the CPU and caches the compressed gradients in CPU memory, thus reducing the GPU memory cost caused by frequent checkpointing.

\begin{figure}[t]
    \centering
    \includegraphics[width=0.5\linewidth]{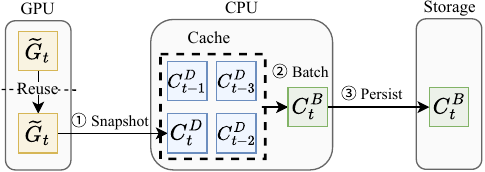}
    \vspace{-5pt}
    \caption{Batched gradient write optimization}
    \Description{Batched gradient write optimization}
    \label{fig:batch}
    \vspace{-20pt}
\end{figure}

\noindent\textbf{Design details. }To solve \textit{Limitation 1 \& 2}, we design a batched gradient writing optimization, which offloads multiple compressed gradients from GPU to CPU memory, and then aggregates them into batched writes to storage. As shown in Figure \ref{fig:batch}, we specify the batched gradient writing optimization of \sysname in three steps:

Step \ding{172}: Offloading gradients to CPU memory. After the training process transmits compressed gradients $\widetilde{G}_t$, the checkpointing process retrieves the GPU memory handles and offloads the gradients to CPU memory. After offloading, the checkpointing process closes the handle of $\widetilde{G}_t$ and frees the corresponding GPU memory.

Step \ding{173}: Batching gradients in buffer. The compressed gradients, which are reused to act as differential checkpoints $C^D_t$, are buffered in CPU memory, awaiting batching. Once the batching size is reached (e.g., the size is four in Fig \ref{fig:batch}), the checkpointing process groups the buffered differential checkpoints $C^D_t$,  $C^D_{t-1}$, $C^D_{t-2}$, and $C^D_{t-3}$ into a batch checkpoint $C^B_t$. 

Step \ding{174}: Writing batched checkpoint to storage. The checkpointing process extracts the batched differential checkpoint $C^B_t$ from the CPU buffer and writes it to storage in a single I/O operation.

\noindent\textbf{Implementation.} Differential checkpoints are offloaded to CPU memory with \texttt{Snapshot()} (non-blocking) and temporarily cached in a buffer. After that, GPU IPC handles are released. Once reaching the batch size, checkpoints are aggregated (tensor addition or dictionary accumulation) and persisted to local or remote storage.

\noindent\textbf{Configuration.} In practice, \sysname selects the batching size $b$ and the full-checkpoint interval $f$ with lightweight rules: $b$ is set so that the buffered differentials can be drained within one training iteration without overflowing the CPU buffer, and $f$ follows the classical MTBF-based checkpoint interval analysis~\cite{daly2006higher} to balance the steady-state checkpointing overhead against the recovery cost.

\section{Enhancement}
\label{sec:enhance}

\sysname's applicability depends on the availability of gradient compression, which, as noted in \S\ref{subsec:dnn}, is not always employed: accuracy-sensitive workloads may avoid lossy compression due to its impact on convergence and model quality, and high-bandwidth interconnects diminish its communication benefits. An important question is therefore whether gradient reuse can be extended to training systems without gradient compression.

In this section, we first outline two key insights regarding snapshotting and persistence mechanisms in the absence of compression. We then present \sysnameplus, an enhanced version of \sysname tailored for scenarios where gradient compression is unavailable or undesired.

\subsection{Issues of Snapshotting and Persistence}
\label{subsec:issue}

\begin{figure}[t]
    \centering
    \subfloat[Issue 1: Snapshotting uncompressed gradient causes additional iteration time\label{fig:snapshot_issue}]{
        \includegraphics[width=0.3\linewidth]{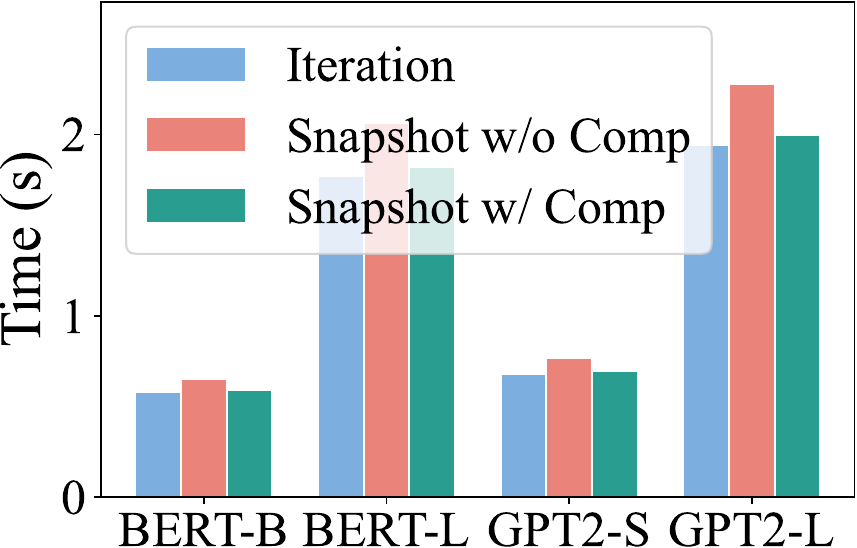}
    }
    \hspace{0.05\linewidth}
    \subfloat[Issue 2: Persisting large volume batched gradients requires multiple iteration time\label{fig:persist_issue}]{
        \includegraphics[width=0.3\linewidth]{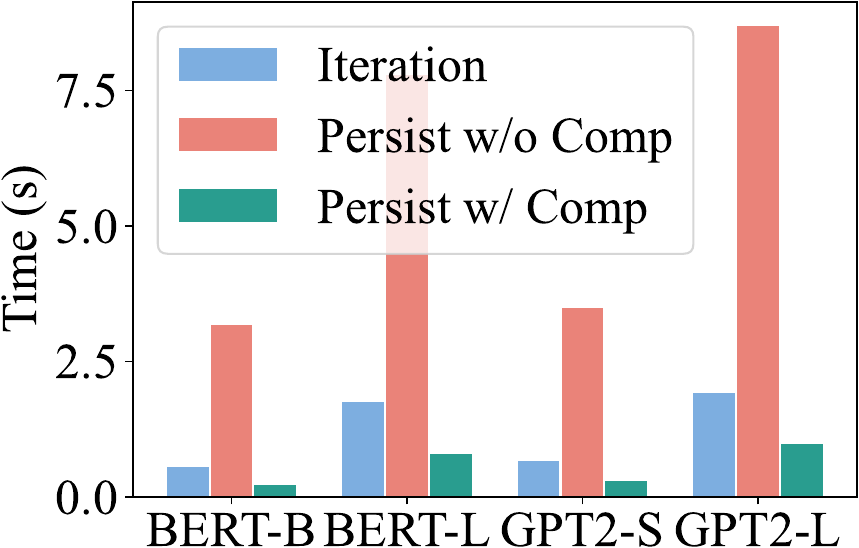}
    }
    \vspace{-10pt}
    \caption{Issues of snapshotting and persistence.}
    \Description{Issues of snapshotting and persistence.}
    \label{fig:issue}
    \vspace{-20pt}
\end{figure}

\noindent\textbf{Snapshotting issue.} A key challenge in the non-compression setting arises during the snapshot phase, where transferring uncompressed gradients from GPU to CPU introduces significantly heavier data movement compared to their compressed counterparts. This increased data traffic intensifies resource contention for GPU and I/O bandwidth and directly competes with training computation, which substantially raises the risk of training stalls.

To quantify the impact of bandwidth contention, we measure the iteration time with and without snapshotting of uncompressed gradients, following the experimental setup in \S\ref{subsec:observations}. As shown in Figure~\ref{fig:snapshot_issue}, snapshotting uncompressed gradients increases the iteration time by 12.3\%, 16.7\%, 13.1\%, and 17.4\% for BERT-B, BERT-L, GPT2-S, and GPT2-L, respectively. In contrast, snapshotting compressed gradients incurs only a 2.4\%–3.1\% overhead.

\noindent\textbf{Persistence issue.}
Although the layer-wise gradient reuse strategy mitigates the snapshotting overhead, a critical issue remains in the persistence phase when gradients are not compressed. The batched persistence strategy introduced in \S\ref{subsec:LowDiff-b} accumulates small compressed updates before writing them to storage, which is effective when updates are sparse and compact. However, without compression, differential checkpoints are substantially larger, and simply applying the same batching mechanism still demands write throughput that can easily saturate available storage bandwidth.

To quantify the persistence overhead, we measure the time required to write a batched gradient checkpoint with and without compression, using the same setup as in \S\ref{subsec:observations} and a fixed batching size of $BS=2$. As shown in Figure~\ref{fig:persist_issue}, persisting uncompressed batched gradients requires 5.51$\times$, 4.41$\times$, 5.14$\times$, and 4.48$\times$ the iteration time for BERT-B, BERT-L, GPT2-S, and GPT2-L, respectively, incurring significant overhead over \sysname's persisting compressed batches (\S\ref{subsec:LowDiff-b}), which can be fully overlapped within a single iteration. Consequently, the volume of uncompressed data becomes the primary bottleneck for persistence.

\begin{figure}[t]
    \centering
    \subfloat[Insight 1: Layer-wise gradient reuse\label{fig:backward}]{
        \includegraphics[width=0.4\linewidth]{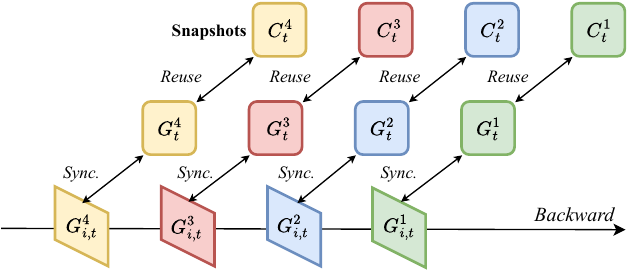}
    }%
    \hspace{0.01\linewidth}%
    \subfloat[Insight 2: Batching strategies comparison\label{fig:model}]{
        \includegraphics[width=0.51\linewidth]{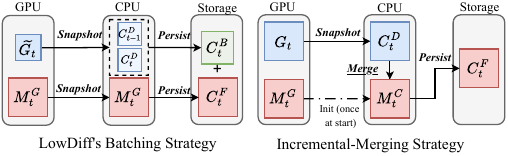}
    }
    \vspace{-10pt}
    \caption{Insights into gradient reuse and batching strategies.}
    \Description{Layer-wise gradient reuse and comparison of batching strategies.}
    \label{fig:insights}
    \vspace{-15pt}
\end{figure}

\subsection{Insights into Snapshotting and Persistence}
\noindent\textbf{Insight 1: Gradient can be reused layer-by-layer.} 
We find that gradients are produced incrementally in the backward pass, following the reverse layer order during training \cite{rumelhart1986learning}. Instead of waiting until the entire backward pass finishes to synchronize all gradients, which would incur high latency and risk blocking training, modern frameworks (e.g., DeepSpeed~\cite{deepspeed}, PyTorch DDP~\cite{li2020pytorch}, and Horovod~\cite{sergeev2018horovod}) already exploit layer-wise gradient availability by initiating communication as soon as each layer’s gradients are produced. Consequently, once a layer’s gradient has been synchronized, we can immediately reuse it in the reusing queue and snapshot it to host memory. This
insight motivates a layer-wise fine-grain reuse and snapshotting design.

As shown in Figure~\ref{fig:backward}, taking the 4-th layer as an example, once its local gradient $G^4_{i,t}$ is computed (Equation~(\ref{eqn:backward})) and synchronized (Equation~(\ref{eqn:sync})), the resulting global gradient $G_t^4$ is immediately enqueued for reuse and snapshotting. Crucially, this process forms a cascading pipeline: as the backward pass propagates to earlier layers (3, 2, and 1), their computation and synchronization overlap with the snapshotting of later layers. This fine-grained interleaving maximizes memory bandwidth utilization and effectively hides the checkpointing overhead within the training execution, thereby minimizing stalls.

\noindent\textbf{Insight 2: Differential and full checkpoints can be merged incrementally.}
We observe that in \sysname, differential checkpoints (i.e., raw gradients) are currently persisted \textit{separately} from full checkpoints. Instead, they can be \textit{incrementally merged} into the CPU-resident full checkpoint following the model update formula (Equation~(\ref{eqn:update})), allowing the CPU to maintain an up-to-date replica of the GPU model. This merging strategy significantly reduces the volume of data to be persisted, as only the full model state, rather than individual differentials, needs to be persisted. This insight motivates an incremental merging design. 

As illustrated in Figure~\ref{fig:model}, \sysname persists both differential ($C^B_t$) and full ($C^F_t$) checkpoints into storage. In contrast, we can maintain a CPU-resident model replica ($M_t^C$). The model state in CPU ($M_t^C$) is initialized only once from the full model state in GPU ($M_t^G$) at the start of training. During subsequent iterations, incoming differential checkpoints ($C^D_t$) are immediately applied to update $M_t^C$ in memory, keeping it synchronized with the GPU model. Consequently, only the full model state in CPU ($C^F_t$) requires persistence. By merging the differentials into the CPU-resident state incrementally, this insight eliminates the overhead of writing separate differential checkpoints to storage.

\subsection{Design of \sysnameplus}

\begin{figure*}[t]
    \centering
    \includegraphics[width=0.95\textwidth]{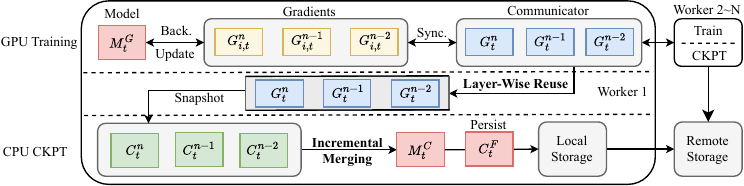}
    \vspace{-10pt}
    \caption{Architecture of \sysnameplus.}
    \Description{Architecture of \sysnameplus.}
    \label{fig:without}
\end{figure*}

We design \sysnameplus for training scenarios without gradient compression. Figure~\ref{fig:without} shows its architecture, and Algorithm~\ref{alg:epicplus} presents the detailed workflow.

\noindent\textbf{Training Process.} The model performs backpropagation to obtain gradients $G^n_{i,t}$ layer-by-layer (Line 4). Following computation, each layer's gradient undergoes synchronization (Line 5). Instead of blocking, the synchronized gradient $G^n_t$ is immediately enqueued into the reusing queue $Q$ (Line 6). This non-blocking operation allows the GPU to proceed with the backward pass of the next layer while the checkpointing process handles the data asynchronously. Finally, the training process aggregates the full gradient $G_t$ (Line 7) to update the GPU model state $M^G_{t+1}$ (Line 8).

\noindent\textbf{Checkpointing Process.} The checkpointing process is structured into two stages: layer-wise-reuse snapshotting (based on \textit{Insight 1}) and incremental-merging persistence (based on \textit{Insight 2}). For layer-wise snapshotting, the checkpointing process continuously fetches gradients $G^n_t$ from the reuse queue $Q$ (Line 11) and performs an immediate snapshot to host memory (Line 12). For incremental-merging persistence, we maintain a CPU-resident replica $M^C_t$. After all layers' gradients are snapshotted, the full gradient is aggregated in the CPU memory (Line 13). Once aggregated, the reused gradients are applied to $M^C_t$ to update its model state (Line 14). The updated model state serves as the in-memory checkpoint. Since this update executes asynchronously alongside the next training iteration in the GPU, its latency is effectively hidden. Finally, persistence is performed asynchronously by writing the updated CPU model replica to storage (Line 15).

\begin{algorithm}[t]
    \caption{\sysnameplus with an uncompressed training process}
    \label{alg:epicplus}
    \small
    \begin{algorithmic}[1]
    \Require GPU model state $M^G_t$ at the $t$-th iteration, gradient $G^n_{i,t}$ of the $n$-th layer on node $i$ , reusing queue $Q$, CPU model state $M^C_t$.
    \end{algorithmic}
    \vspace{2pt}
    \noindent\begin{minipage}[t]{0.49\linewidth}%
    \begin{algorithmic}[1]
    \Function{Training Process}{}
        \ForAll{$t$}
            \ForAll{$n$} \Comment{Layer-by-Layer}
                \State $G^n_{i,t} \gets Backward(x^n_t)$ \Comment{Backward}
                \State $G^n_t \gets Sync(G^n_{i,t})$ \Comment{Synchronization}
                \State $\textcolor{LightBlue}{\textbf{Q.put}}(G^n_t)$ \Comment{Gradient Reuse}
            \EndFor
            \State $G_t \gets \sum G^n_t $  \Comment{Wait for Sync}
            \State $M^G_{t+1} \gets M^G_t + Adam(o_t, G_t)$ \Comment{Update}
        \EndFor
    \EndFunction
    \end{algorithmic}
    \end{minipage}\hfill
    \begin{minipage}[t]{0.49\linewidth}%
    \begin{algorithmic}[1]
    \algsetline{8}
    \Function{Checkpointing Process}{}
        \ForAll{$t$}
            \While{$G^n_t \gets \textcolor{LightBlue}{\textbf{Q.get}}$} \Comment{Gradient Reuse}
                \State $C^n_t \gets Snapshot(G^n_t)$ \Comment{Snapshot}
            \EndWhile
            \State $G_t \gets \sum C^n_t$  \Comment{Wait for Snapshot}
            \State $M^C_{t+1} \gets M^C_t + Adam(o_t, G_t)$ \Comment{Merge}
            \State $C^F_t \gets Persist(M^C_t)$ \Comment{Persist Regularly}
        \EndFor
    \EndFunction
    \end{algorithmic}
    \end{minipage}
\end{algorithm}

\noindent\textbf{Implementation.}
Both off-loaded operations are scheduled according to their contended resources. For layer-wise-reuse snapshotting, \texttt{backward\_hooks} trigger the snapshotting immediately upon each layer's gradient synchronization during the backward pass, and a \texttt{ThreadPoolExecutor} performs the asynchronous D2H transfers. Since these transfers compete for PCIe and host-memory bandwidth, they are issued on streams prioritized below the gradient-synchronization streams: under bandwidth contention they yield to training communication, so the bulk of the transfers is absorbed during the interconnect-idle phases (the model update and the following forward propagation). The transfers are sharded across GPU ranks to exploit parallel DMA paths, with large tensors copied in continuous chunks into pre-allocated pinned host buffers and small tensors packed into fixed-size contiguous blocks. For incremental-merging persistence, the CPU-resident replica is initialized once via \texttt{copy.deepcopy()} and partitioned into balanced shards that are updated in parallel by multiple workers using DeepSpeed CPUAdam with vectorized kernels (e.g., AVX-512); these CPU-side updates occupy CPU cores and host memory bandwidth and are therefore executed during backward propagation, where GPU computation and gradient synchronization dominate the iteration time and the host is relatively idle. Storage persistence is likewise asynchronous and off the training critical path. For recovery, the checkpointing process is isolated from the training process: a software failure only requires reinitializing the training process and restoring the CPU-resident states directly to GPUs without disk access, whereas a hardware failure recovers from the latest persisted checkpoint, similar to prior methods~\cite{mohan2021checkfreq}.

\section{Evaluation}
\label{sec:evaluation}
\subsection{Experimental Setup}
\label{subsec:setup}

\noindent\textbf{Servers.}
We evaluate \sysname and \sysnameplus\footnote{Open-sourced at \url{https://github.com/YuchongHu/LowDiff}} on A100 and V100S servers, equipped with four NVIDIA A100 (80GB) or V100S (32GB) GPUs and Intel Xeon 8352V or 4214 CPUs, respectively.
Each server has 512GB DDR4 DRAM and a 4TB Samsung SSD.
GPUs communicate via NVLink within servers and 25Gbps Mellanox ConnectX-5 InfiniBand across servers.
All servers run Ubuntu 22.04 with DeepSpeed 0.16.4, CUDA 12.4, and PyTorch 2.6.0.

\begin{table*}[t]
    \centering
    \small
    \setlength{\tabcolsep}{4pt}
    \begin{tabular}{lcccccccc}
        \toprule
        & ResNet-101 & VGG-19 & BERT-B & BERT-L
        & GPT2-S & GPT2-L & Qwen3-1.7B & Qwen3-4B \\
        \midrule
        Dataset
        & ImageNet & ImageNet & SQuAD & SQuAD
        & WikiText & WikiText & RedPajama & RedPajama \\
        
        Params
        & 44.5M & 143M & 110M & 334M
        & 117M & 762M & 1.70B & 4.09B \\
        \bottomrule
    \end{tabular}
    \vspace{5pt}
    \caption{Models and datasets used for evaluation.}
    \label{tab:setup}
    \vspace{-20pt}
\end{table*}

\noindent\textbf{Workloads.}
We evaluate image classification and NLP models, including ResNet~\cite{he2016deep}, VGG~\cite{simonyan2014very}, BERT~\cite{devlin2018bert}, GPT2~\cite{radford2019language}, and Qwen3~\cite{qwen3}, on CIFAR-100~\cite{krizhevsky2009learning}, ImageNet~\cite{deng2009imagenet}, SQuAD~\cite{SQuAD}, WikiText~\cite{merity2016pointer}, and RedPajama-Data-1T~\cite{weber2024redpajamaopendatasettraining}. Table~\ref{tab:setup} summarizes the workload configurations. We use Adam~\cite{kingma2014adam} with a sparsification ratio of $\rho=0.01$~\cite{aji2017sparse,li2022near,fei2021efficient,ming2024adtopk,chen2023bbtopk}.

\noindent\textbf{Comparative Methods.}
We compare against CheckFreq~\cite{mohan2021checkfreq}, Gemini~\cite{wang2023gemini}, and Naïve DC, using their default configurations. CheckFreq asynchronously persists full checkpoints, while Gemini maintains snapshots in CPU memory. Naïve DC adapts the differential checkpointing scheme of Check-N-Run~\cite{eisenman2022check} to general DNN training, compresses model differentials and persists them, leaving optimizer states uncompressed. Similarly, \sysname provides per-iteration durable checkpoints by reusing compressed gradients with batched writes. \sysnameplus separates per-iteration CPU snapshots (\sysnamepluss) for software-failure recovery from periodic storage persistence (\sysnameplusp) for hardware-failure recovery.

\subsection{Performance Evaluation}
\label{subsec:performance}

\noindent\textbf{Training time with gradient compression (Exp.~1).} We evaluate the training time with gradient compression of \sysname and \sysnameplus under different checkpointing strategies on the previously introduced DNN models, with the checkpointing frequency set to once per iteration. For each model, we run 1,000 training iterations on A100 servers and record the total training time. The \textit{W/O CKPT} setting represents training without checkpointing, which serves as the upper bound of training speed.

Figure \ref{fig:training_time} shows that the training time of \sysname and \sysnameplus on all eight training tasks is close to that of W/O CKPT and significantly shorter than that of other solutions. Compared to W/O CKPT, \sysname and \sysnameplus's training time increases by only 2.4\%-3.1\% and 0.7\%-2.9\%, respectively, while the other checkpointing methods' training time increases by 8.1\%-891\%.

Specifically, we elaborate on \sysname and \sysnameplus's benefits over the other methods and analyze the reasons behind these benefits:

\begin{figure*}[t]
    \setlength{\abovecaptionskip}{2pt}
    \centering
    \includegraphics[width=\textwidth]{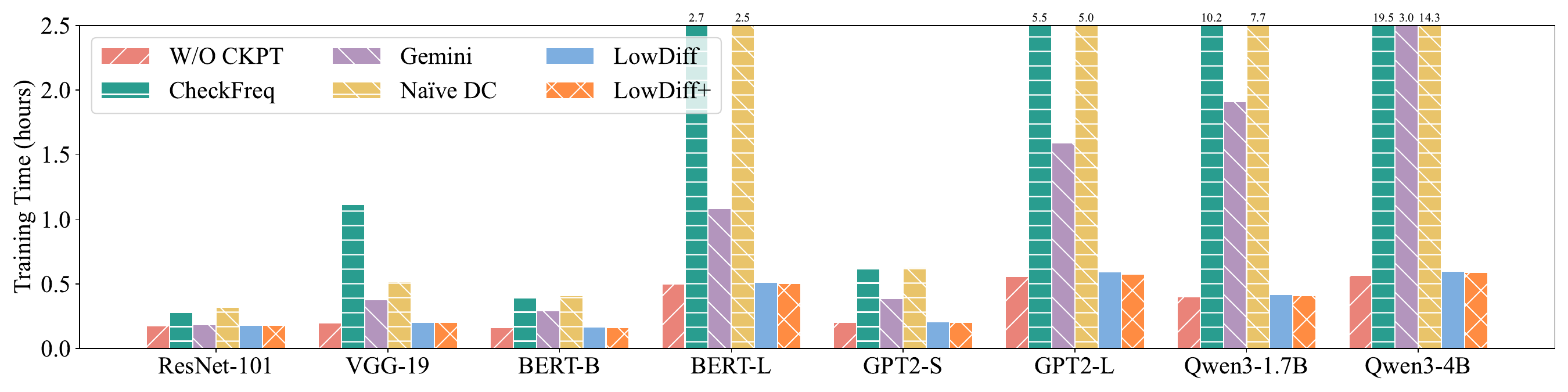}
    \caption{(Exp.~1) Training time with gradient compression.}
    \Description{Training time with gradient compression.}
    \label{fig:training_time}
    \vspace{-10pt}
\end{figure*}

\begin{itemize}[leftmargin=10pt]
  \item Compared to CheckFreq and Gemini in GPT2-S, \sysname reduces the training time by 68.2\% and 46.1\%, respectively. We see that \sysname significantly outperforms them under high frequency since \sysname uses differential checkpointing with compression (compression ratio=0.01) to largely reduce the checkpoint sizes. Further, Gemini improves its training performance over CheckFreq by checkpointing to CPU memory, but the improvement is limited by its large checkpoint size.

  \item \sysname's performance improvement is more pronounced for larger models like BERT-L, GPT2-L, and Qwen3-4B. Compared to CheckFreq and Gemini in GPT2-L, \sysname reduces the training time by 89.2\% and 59.2\%, respectively. For Qwen3-4B, the reduction increases to 94.1\% and 80.8\%, respectively. This is because, as the model size increases, the advantage of the reduced checkpoint size of \sysname becomes more dominant on the overall training time.

  \item Compared to Naïve DC in BERT-B, \sysname reduces the training time by 60.5\% due to its unique gradient reusing scheme, which reduces and even eliminates compression and transmission cost of Naïve DC, confirming the efficiency of our reusing idea in \S\ref{subsec:motivation}. It should be noted that, in comparison with CheckFreq and Gemini, Naïve DC fails to demonstrate the advantage of its reduced checkpoint size due to its high costs.

  \item \sysnameplus is even faster than \sysname, reducing the training time by 0.8\%--3.3\% across models. The gain is limited because compressed gradients are already small enough for \sysname's DC; the advantages of \sysnameplus become pronounced in non-compression scenarios (Exp.~2).
\end{itemize}

\noindent\textbf{Training time without gradient compression (Exp.~2).} To demonstrate the generality of gradient reuse beyond compression-enabled training, we evaluate \sysname and \sysnameplus without gradient compression. In this setting, we disable gradient compression and enable NVLink-based GPU communication to mitigate the communication overhead caused by transmitting full gradients. All other configurations remain the same as Exp.~1.

\begin{figure*}[t]
    \includegraphics[width=\textwidth]{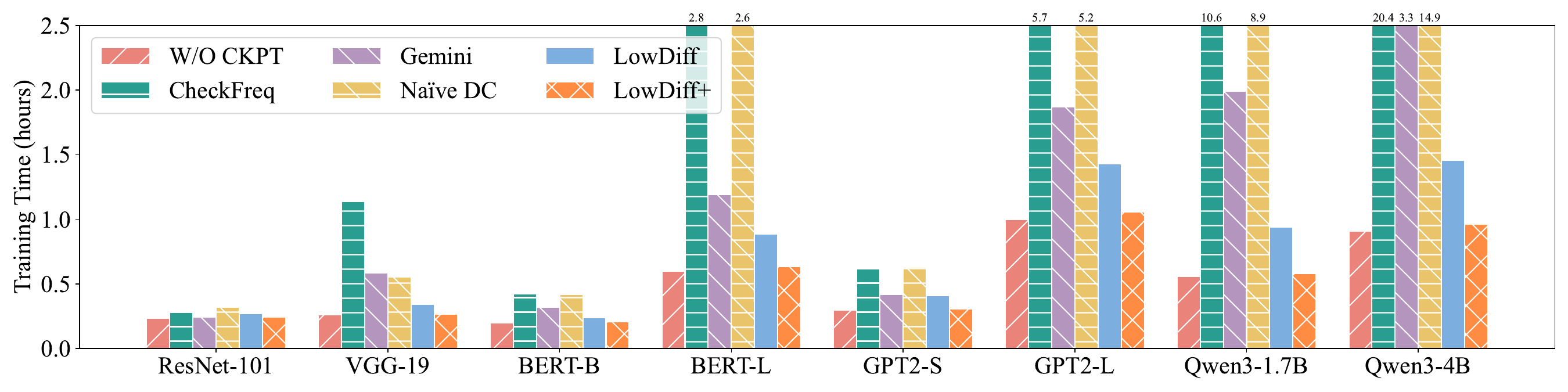}
    \vspace{-20pt}
    \caption{(Exp.~2) Training time without gradient compression.}
    \Description{Training time without gradient compression.}
    \label{fig:training_time+}
    \vspace{-10pt}
\end{figure*}

As shown in Figure~\ref{fig:training_time+}, \sysnameplus achieves the lowest training time among all evaluated methods. For the GPT2-L training task, it reduces the training time by 42.4\% compared to Gemini and by 81.2\% compared to CheckFreq. For the Qwen3-4B training task, \sysnameplus further reduces the training time by 81.9\% compared to Gemini and by 96.7\% compared to CheckFreq, demonstrating its effectiveness for large-scale training without gradient compression. This efficiency stems from gradient reuse: the reused gradients are only one-third the size of full training states (Finding~2), and layer-wise snapshotting and incremental merging hide the remaining costs (\S\ref{sec:enhance}).

Table~\ref{tab:lowdiffplus_breakdown} provides an iteration-level execution breakdown of \sysnameplus. The results show that gradient reuse introduces two additional costs beyond normal training: layer-wise gradient snapshotting (247.3 ms for D2H transfers) and CPU-resident replica updates (442.6 ms for CPUAdam). Following the phase-based scheduling in \S\ref{sec:enhance}, both operations are effectively hidden behind training execution, exposing only 133.9 ms additional overhead on the critical path and limiting the training overhead to 2.9\%--7.2\% compared with W/O CKPT.

\begin{table}[t]
\centering
\begin{tabular}{lcc}
\toprule
\textbf{Component} & \textbf{W/O CKPT} & \textbf{\sysnameplus} \\
\midrule
Forward & 219.1 ms & 224.3 ms \\
Backward+Sync. & 2,475 ms & 2,601 ms \\
Model Update & 165.6 ms & 168.6 ms \\
\midrule
Layer-wise Snapshotting & -- & 247.3 ms \\
CPU-resident Replica Update & -- & 442.6 ms \\
\bottomrule
\end{tabular}
\vspace{5pt}
\caption{(Exp.~2) Iteration-level breakdown of \sysnameplus on Qwen3-4B.}
\label{tab:lowdiffplus_breakdown}
\vspace{-20pt}
\end{table}

\noindent\textbf{Wasted time (Exp.~3).} We evaluate the wasted time of various checkpointing methods in the event of failures. Failures were simulated during training by adhering to a fixed Mean Time Between Failures (MTBF) metric. Using three MTBF values at 0.5, 1, and 2 hours \cite{gupta2024just}, respectively, we evaluate the wasted time of Naïve DC, CheckFreq, Gemini, and \sysname when training GPT2-S. Failures are injected by terminating the training process at timestamps generated according to the specified MTBF. For a software failure, we terminate and restart only the training process while keeping the independent checkpointing process and its CPU-resident state intact. For \sysnameplus, we evaluate its wasted time under both software and hardware failures, where the corresponding checkpoints can be accessed through \sysnamepluss and \sysnameplusp.

\begin{figure*}[t]
    \centering

    \begin{tabular}{@{}c@{\hspace{0.03\textwidth}}c@{}}

        % ==================== Left: Exp. 3 ====================
        \begin{minipage}[t][1.35in][b]{0.48\textwidth}
            \vspace{0pt}
            \centering
            \includegraphics[width=\linewidth]{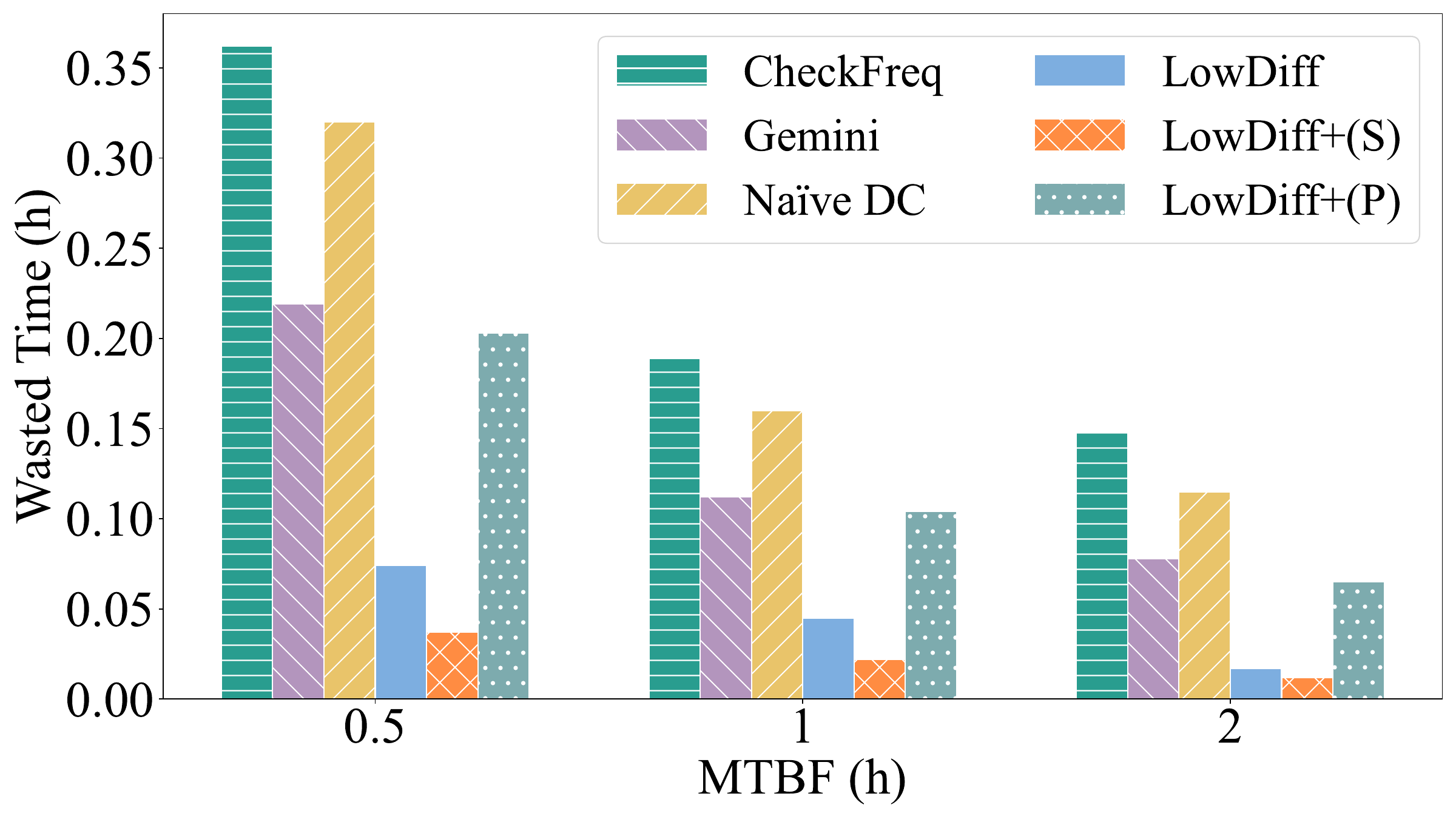}
        \end{minipage}
        &
        % ==================== Right: Exp. 4 ====================
        \begin{minipage}[t][1.35in][b]{0.48\textwidth}
            \vspace{0pt}
            \centering
            \includegraphics[width=\linewidth]{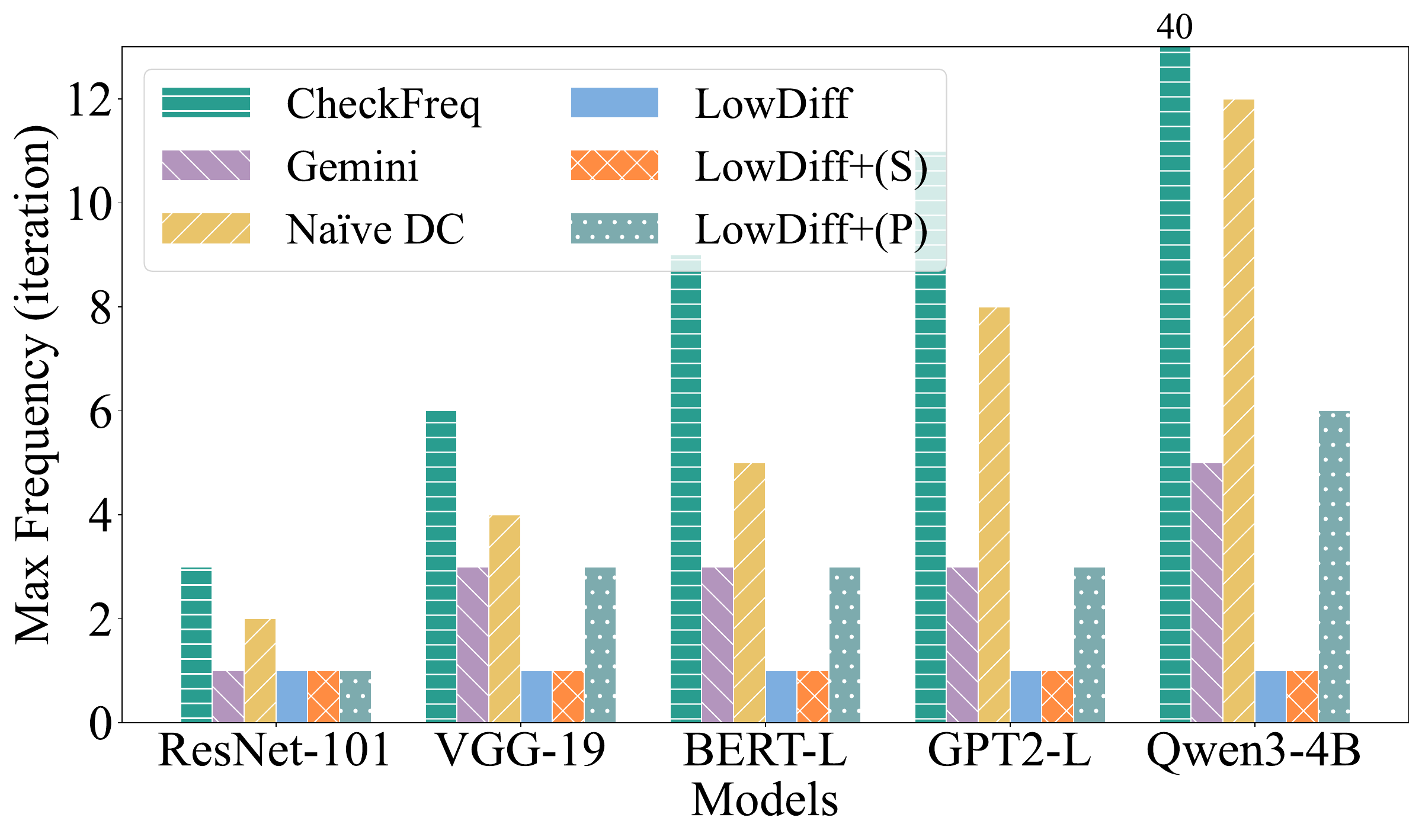}
        \end{minipage}
        \\[-6pt]
        \begin{minipage}[t][0.35in][t]{0.48\textwidth}
            \vspace{0pt}
            \captionsetup{skip=0pt}
            \caption{(Exp.~3) Wasted time under different MTBFs.}
            \Description{Wasted time under different MTBFs.}
            \label{fig:runtime}
        \end{minipage}
        &
        \begin{minipage}[t][0.35in][t]{0.48\textwidth}
            \vspace{0pt}
            \captionsetup{skip=0pt}
            \caption{(Exp.~4) Maximum checkpointing frequency.}
            \Description{Maximum checkpointing frequency.}
            \label{fig:freq}
        \end{minipage}
    \end{tabular}
    \vspace{-20pt}
\end{figure*}

Figure \ref{fig:runtime} shows that the wasted time gap between \sysname and other checkpointing methods keeps increasing as failures become more frequent, while \sysname maintains the lowest wasted time. When the MTBF decreases from 2 to 0.5 (lower MTBF means more frequent failures), the gap between \sysname and Gemini increases from 0.061h to 0.145h. This is because \sysname's reused compressed gradients keep differential checkpoints small, incurring low steady-state checkpointing overhead and fast recovery, while other methods' wasted time is greatly influenced by their checkpoint size and frequency. The results confirm the efficiency of \sysname in the event of failures.

For \sysnameplus, the wasted time due to software failures is 37.1\%–51.4\% lower than that of \sysname. This improvement comes from \sysnameplus's in-memory checkpointing (snapshotting), which enables fast recovery using the full model state stored in CPU memory. The detailed recovery times of \sysnamepluss are presented in Exp.~5. In contrast, the wasted time from hardware failures is 2.74$\times$-3.82$\times$ higher than that of \sysname, due to \sysnameplus’s lower persistence frequency, but still lower than that of CheckFreq and Gemini. Details of the checkpointing frequency for \sysnamepluss and \sysnameplusp are presented in Exp.~4.

\noindent\textbf{Checkpointing frequency (Exp.~4).} To evaluate the maximum checkpointing frequency that each method can achieve under bounded training speed (set as 3.5\%, following Microsoft's study \cite{mohan2021checkfreq}), we measure the minimum checkpoint interval that each method can support while keeping the training slowdown below this bound. Following prior checkpointing studies \cite{eisenman2022check}, we define the checkpoint interval as the number of training iterations between two consecutive checkpoint events. Therefore, for traditional checkpointing methods, the checkpoint events correspond to full checkpoints, while for DC, \sysname, and \sysnamepluss, they correspond to differential checkpoints.

Figure \ref{fig:freq} shows that compared to other methods, \sysname can achieve per-iteration checkpointingfor different widely-used models under the bounded training speed. \sysnamepluss also performs per-iteration in-memory checkpointing, enabled by its efficient layer-wise-reuse snapshotting approach. \sysnameplusp achieves per-iteration persistence frequency in smaller models such as ResNet-101, but the interval increases to 5 iterations for larger models like Qwen3-4B due to the limited bandwidth of SSDs. Nevertheless, its checkpointing intervals remain shorter than those of Naïve DC and CheckFreq, and are nearly the same as those of Gemini.

Although Naïve DC reduces checkpoint size through differential checkpointing, its checkpointing still introduces additional computation and transmission overhead (Challenges 1 and 2). In contrast, Gemini benefits from asynchronous in-memory checkpointing, enabling better overlap with training. Therefore, it achieves shorter checkpoint intervals than Naïve DC despite using full checkpoints (1 iteration for small models and 5 iterations for large models, compared with 2 to 12 iterations for Naïve DC). These results validate our motivation that \sysname and \sysnameplus can support high-frequency checkpointing with minimal overhead.

\noindent\textbf{Recovery time (Exp.~5).} We evaluate the recovery time of different strategies under different full checkpointing frequencies in Qwen3-4B. We compare traditional checkpointing without differential checkpointing as Baseline (i.e., \texttt{torch.save} \cite{paszke2019pytorch}), Na\"ive DC, \sysname, and \sysnamepluss for recovery from software failures. Note that recovery from hardware failures in \sysnameplus is the same as in the baseline, since both reload the model from storage and resume training from the latest persisted checkpoint, as determined by the checkpointing frequency. The recovery time includes reading, decompression, and reconstruction.

\begin{figure*}[t]
    \centering

    \begin{tabular}{@{}c@{\hspace{0.03\textwidth}}c@{}}

        % ==================== Left: Figure ====================
        \begin{minipage}[t][1.35in][c]{0.48\textwidth}
            \vspace{0pt}
            \centering
            \includegraphics[width=\linewidth]{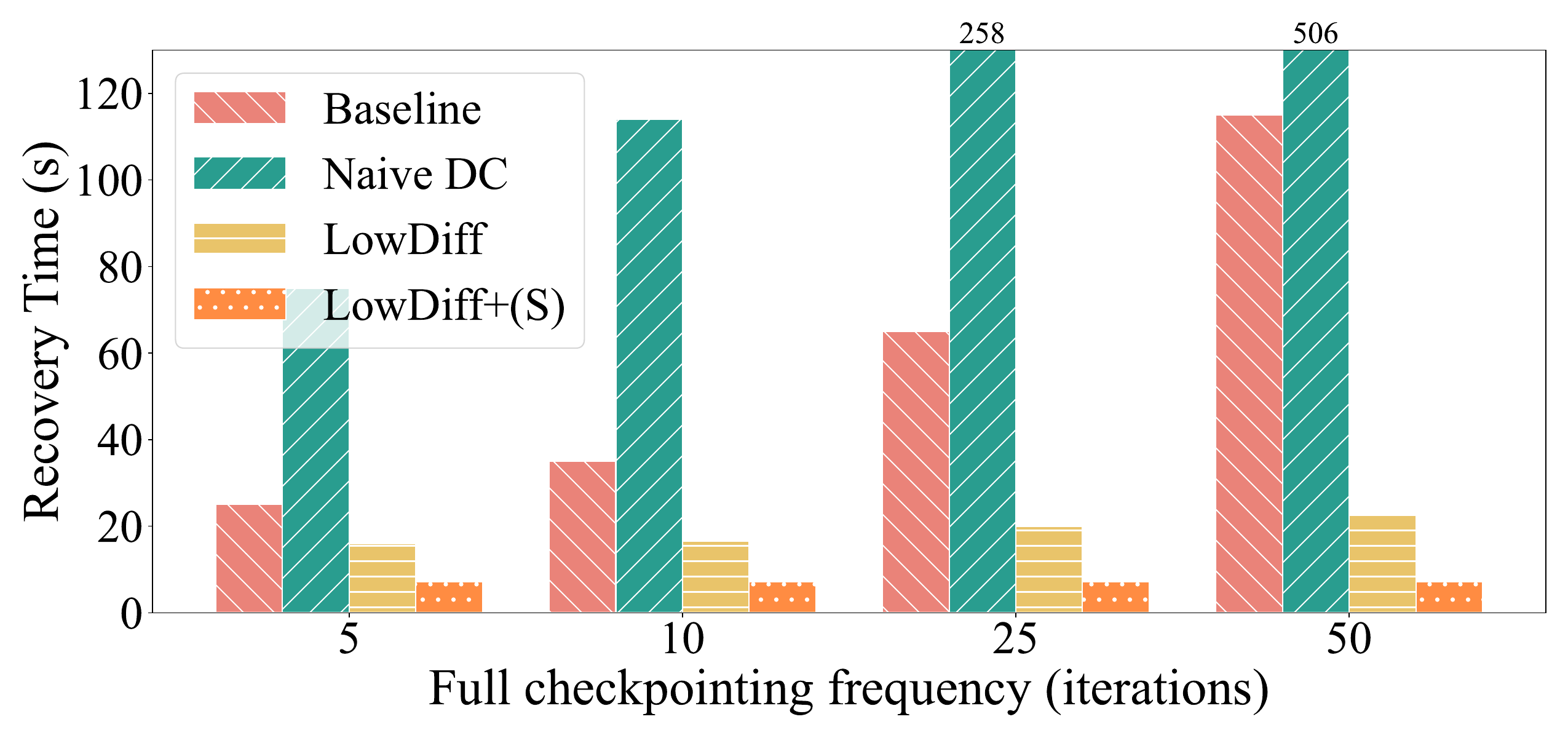}
        \end{minipage}
        &
        % ==================== Right: Table ====================
        \begin{minipage}[t][1.35in][c]{0.48\textwidth}
            \vspace{0pt}
            \centering
            \small
            \setlength{\tabcolsep}{4pt}
            \begin{tabular}{lccc}
                \toprule
                Method & Load & Decomp. & Merge \\
                \midrule
                Baseline          & 15s   & --     & 2.1s   \\
                Naïve DC          & 10s   & 0.138s & 0.086s \\
                \sysname          & 0.14s & 0.139s & 0.093s \\
                \sysnamepluss     & 7.1s  & --     & --      \\
                \bottomrule
            \end{tabular}
        \end{minipage}

        \\[-6pt]

        % ==================== Captions ====================
        \begin{minipage}[t][0.35in][t]{0.48\textwidth}
            \vspace{0pt}
            \captionsetup{skip=0pt}
            \captionof{figure}{(Exp.~5) Recovery time of different methods.}
            \Description{Recovery time of different methods.}
            \label{fig:recovery_time}
        \end{minipage}
        &
        \begin{minipage}[t][0.35in][t]{0.48\textwidth}
            \vspace{0pt}
            \captionsetup{skip=0pt}
            \captionof{table}{(Exp.~5) Recovery breakdown.}
            \label{tab:recovery_breakdown}
        \end{minipage}
    \end{tabular}
    \vspace{-20pt}
\end{figure*}

Figure \ref{fig:recovery_time} shows that the recovery time of \sysname is consistently shorter than that of Baseline and Na\"ive DC. For example, when the full checkpointing frequency is 50 iterations, \sysname reduces the recovery time by 80.4\% and 95.6\% compared with Baseline and Na\"ive DC, respectively. Although a larger full-checkpointing interval requires replaying more differential checkpoints, the recovery time of \sysname grows only sublinearly, since the per-differential costs are small and pipelined with each other (Table~\ref{tab:recovery_breakdown}), leaving the one-time full-checkpoint load as the dominant floor. In contrast, \sysnameplus achieves the shortest recovery time overall, as it performs in-memory differential checkpointing every iteration. By maintaining an in-memory checkpoint state, \sysnameplus avoids persistent storage access during software-failure recovery, which delivers 3.5$\times$–16.2$\times$ faster recovery than Baseline across full checkpointing frequencies from 5 to 50.

The breakdown at the 50-iteration full-CKPT frequency in Table~\ref{tab:recovery_breakdown} further explains the recovery efficiency. The table reports per-record costs rather than end-to-end totals: DC-based methods pay the three-stage cost for each replayed differential, the Baseline pays a one-time full-checkpoint load plus retraining at 2.1 s per iteration (repeating the full forward and backward passes), and \sysnamepluss's 7.1 s is its total one-step restore; aggregating these costs over the lost iterations yields the end-to-end recovery times in Figure~\ref{fig:recovery_time}. We observe that the dominant bottleneck of differential checkpointing is repeated checkpoint I/O rather than differential reconstruction. For Na\"ive DC, checkpoint reading contributes more than 97\% of the total recovery latency, while decompression and reconstruction introduce negligible overhead. These results demonstrate that optimizer states contribute significantly to differential checkpoint size, while recovery computation is not the primary bottleneck. \sysname avoids this bottleneck by reusing compressed gradients instead of persisting optimizer differentials (Finding~2), and its pipelined recovery further overlaps reading, decompression, and reconstruction, achieving the lowest recovery latency among persistent-checkpointing approaches.

\noindent\textbf{Impact of checkpoint/recovery (Exp.~6).} We evaluate whether \sysname and \sysnameplus preserve Qwen3-4B's performance after C/R (i.e., checkpoint/recovery). Note that gradient compression or not is determined by the original training process rather than introduced by \sysname or \sysnameplus. Both approaches only reuse gradients already generated during training to construct differential checkpoints. We first compare the perplexity of recovered models with the corresponding pre-failure models (baseline) under identical training configurations. We further analyze the state-level and output-level differences to quantify the C/R fidelity.

Table~\ref{tab:model_correctness} first reports the perplexity of the recovered models. Both \sysname and \sysnameplus recover the same perplexity as their corresponding pre-failure models, demonstrating that gradient reuse does not introduce observable degradation in model performance. The slightly higher perplexity of the model trained with gradient compression is caused by the original gradient compression configuration rather than C/R.

\begin{table}[t]
\centering
\begin{tabular}{lccccc}
\toprule
Method & PPL & $\Delta$Model & $\Delta$Optim. & KL & Top-10 \\
\midrule
Baseline w/ GC   & 15.70 & 0 & 0 &  0 & 100\% \\
\sysname    & 15.70 & 0 & 0 & 0 & 100\% \\
\midrule
\midrule
Baseline w/o GC  & 15.42 & 0 & 0 & 0  & 100\% \\
\sysnameplus & 15.42 & 1e-9 & 2e-9 &  1e-10 & 100\% \\
\bottomrule
\end{tabular}
\vspace{5pt}
\caption{(Exp.~6) Model performance after checkpoint/recovery. GC denotes gradient compression.}
\vspace{-20pt}
\label{tab:model_correctness}
\end{table}

We further analyze C/R fidelity by comparing the recovered states and outputs with the corresponding pre-failure models. \sysname achieves identical recovery, with zero differences in model parameters and optimizer states. For \sysnameplus, CPU-side optimizer updates introduce negligible floating-point variations due to different execution paths between CPU and GPU, with state differences remaining at the $\mathcal{O}(10^{-9})$ scale. These variations do not affect model behavior: \sysnameplus introduces only a $1\times10^{-10}$ KL divergence and achieves 100\% top-10 token prediction consistency. These results demonstrate that gradient reuse preserves both recovery fidelity and model performance after C/R.

\begin{figure}[t]
    \centering
    \subfloat[Average checkpointing time under different batch sizes\label{fig:batching}]{
        \includegraphics[width=0.3\textwidth]{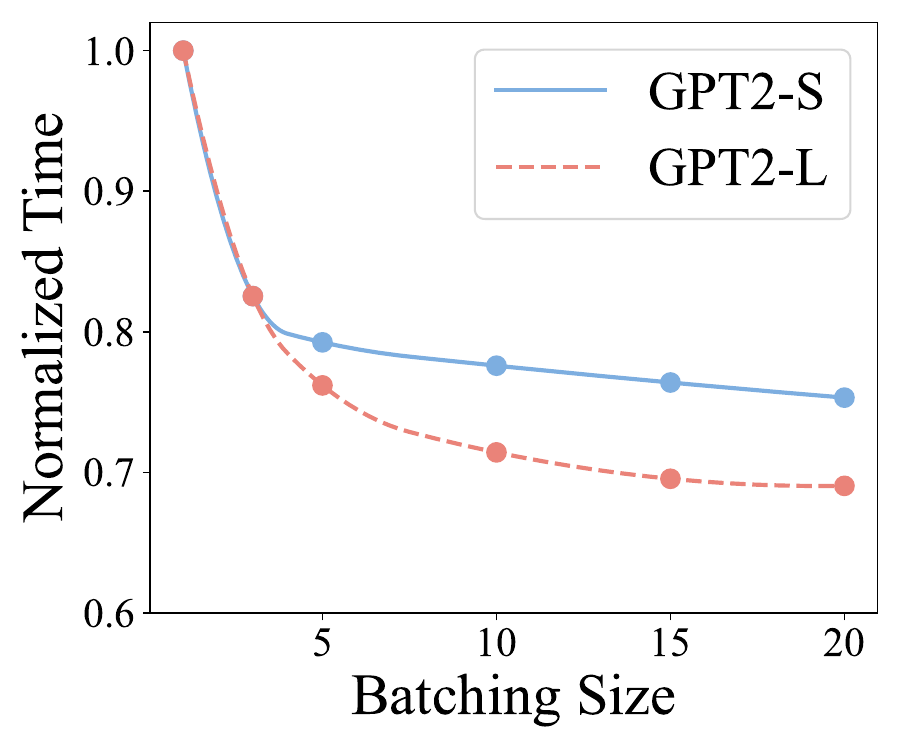}
    }
    \hspace{0.05\linewidth}
    \subfloat[GPU memory consumption w/o and w/ offloaded batching\label{fig:memory}]{
        \includegraphics[width=0.3\textwidth]{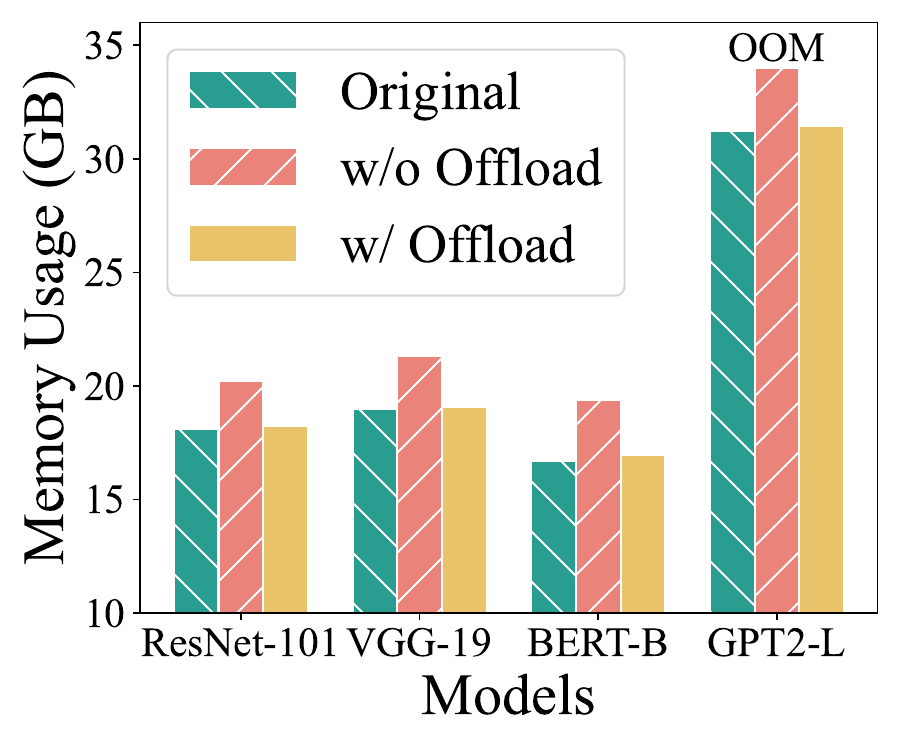}
    }
    \vspace{-8pt}
    \caption{(Exp.~7) CKPT time and GPU memory reduction.}
    \Description{CKPT time and GPU memory reduction.}
    \vspace{-20pt}
\end{figure}

\noindent\textbf{Checkpointing time reduction and GPU memory reduction (Exp.~7).} We evaluate the reduction in average checkpointing time achieved by the batched write optimization proposed in \S\ref{subsec:LowDiff-b} on various models with different batching sizes. Figure \ref{fig:batching} shows that the batched write strategy significantly reduces the average time required to write differential checkpoints, with a maximum reduction of $30.9\%$ when the batching size is 20 during GPT2-S training. 

The batching strategy effectively accelerates the checkpoint writing process and adapts to different compression ratios, minimizing the impact on training performance.

Additionally, we also evaluate the reduction of GPU memory achieved by the batching offloaded to CPU (\S\ref{subsec:LowDiff-b}). Figure \ref{fig:memory} shows that the GPU memory usage will increase by $10\%-12\%$ without offloaded batching, especially for GPT2-L, potentially leading to memory exhaustion. In contrast, the GPU memory usage will return to the original state with the help of offloaded batching. 

\begin{figure*}[t]
    \centering

    \begin{tabular}{@{}c@{\hspace{0.03\textwidth}}c@{}}

        % ==================== Body ====================
        \begin{minipage}[t][1.20in][c]{0.46\textwidth}
            \vspace{0pt}
            \centering
            \small
            \setlength{\tabcolsep}{4pt}
            \begin{tabular}{lccc}
                \toprule
                & VGG-19 & BERT-L & GPT2-L \\
                \midrule
                Full CKPT & 1.7G  & 3.8G  & 8.7G \\
                Naïve DC  & 1.13G & 2.55G & 5.7G \\
                \sysname   & 109M  & 239M  & 541M \\
                \bottomrule
            \end{tabular}
        \end{minipage}
        &
        \begin{minipage}[t][1.20in][c]{0.46\textwidth}
            \vspace{0pt}
            \centering
            \includegraphics[width=\linewidth]{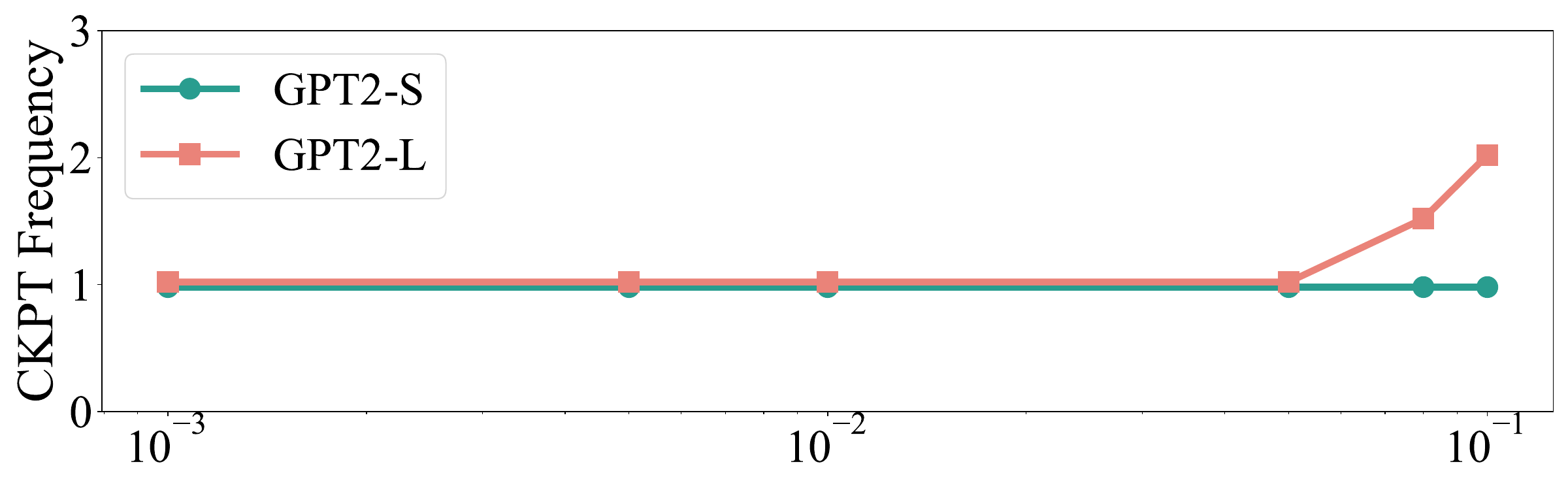}
        \end{minipage}
        \\[-12pt]

        \begin{minipage}[t][0.35in][t]{0.46\textwidth}
            \vspace{0pt}
            \captionsetup{skip=0pt}
            \captionof{table}{(Exp.~8) Storage overhead of checkpoints.}
            \label{tab:storage}
        \end{minipage}
        &
        \begin{minipage}[t][0.35in][t]{0.46\textwidth}
            \vspace{0pt}
            \captionsetup{skip=0pt}
            \captionof{figure}{(Exp.~9) Impact of compression ratio $\rho$.}
            \Description{Impact of compression ratio on checkpointing performance.}
            \label{fig:ratio}
        \end{minipage}
    \end{tabular}
    \vspace{-20pt}
\end{figure*}

\noindent\textbf{Storage overhead (Exp.~8).}
We evaluate the storage overhead of checkpoints for various checkpointing methods and various DNN models, where CheckFreq and Gemini only have full checkpoints (called Full CKPT), Naïve DC and \sysname have full and differential checkpoints. As shown in Table~\ref{tab:storage}, compared to full checkpoints, Naïve DC reduces the storage overhead by 34.4\% due to the compression technique used in the model parameters. 

Compared to Naïve DC, \sysname further reduces the storage overhead by 90.5\%. Because \sysname reuses compressed gradients to construct DCs for both model parameters and optimizer parameters (Finding 2 in \S\ref{subsec:insight}), while Naïve DC (Check-N-Run \cite{eisenman2022check}) does not apply sparsification compression to the optimizer parameters within DCs, resulting in the optimizer occupying the majority of the checkpoint space. \sysnameplus does not introduce any compression and thus exhibits the same storage overhead as baseline. The results confirm our motivation that \sysname can significantly reduce the checkpoint size and thus lower the storage cost.

\noindent\textbf{Impact of compression ratio $\rho$ (Exp.~9).}
We evaluate the impact of different compression ratios $\rho$ on the checkpoint frequency of GPT2-L and GPT2-S training. The compression ratio $\rho$ varies from 0.1 to 0.001, consistent with the ranges commonly used in prior work \cite{li2022near, ming2024adtopk}. 

As shown in Figure \ref{fig:ratio}, \sysname enables per-iteration checkpointing for GPT2-S training across the entire range of $\rho \in [0.001, 0.1]$ due to its smaller enough gradient size for checkpointing to be overlapped with training within one iteration. For GPT2-L, \sysname supports per-iteration checkpointing when $\rho \in [0.001, 0.075]$, while at $\rho = 0.1$, the checkpoint frequency increases to 2 iterations because the increasing gradient size makes the checkpointing take two iterations to be overlapped with training. The result shows \sysname ensures frequent checkpointing (intervals \textless 3 iterations) under the common compression ratio.

\subsection{System Scalability}
We test the scalability of \sysname and \sysnameplus on V100 servers in scenarios with frequent software failures and more GPUs.

\noindent\textbf{Scaling to frequent software failures (Exp.~10).} We evaluate all methods under more frequent software failures, with MTBF ranging from 0.1 to 5 hours. Using the effective training time ratio metric as defined in Gemini \cite{wang2023gemini}, which measures the percentage of time spent on productive training, we report the ratio of \texttt{torch.save} (baseline), CheckFreq, Gemini, \sysname, and \sysnameplus.

\begin{figure*}[t]
    \centering
    \begin{tabular}{@{}c@{\hspace{0.03\textwidth}}c@{}}
        % ==================== Left: Exp. 10 ====================
        \begin{minipage}[t][1.55in][c]{0.48\textwidth}
            \vspace{0pt}
            \centering
            \includegraphics[width=\linewidth]{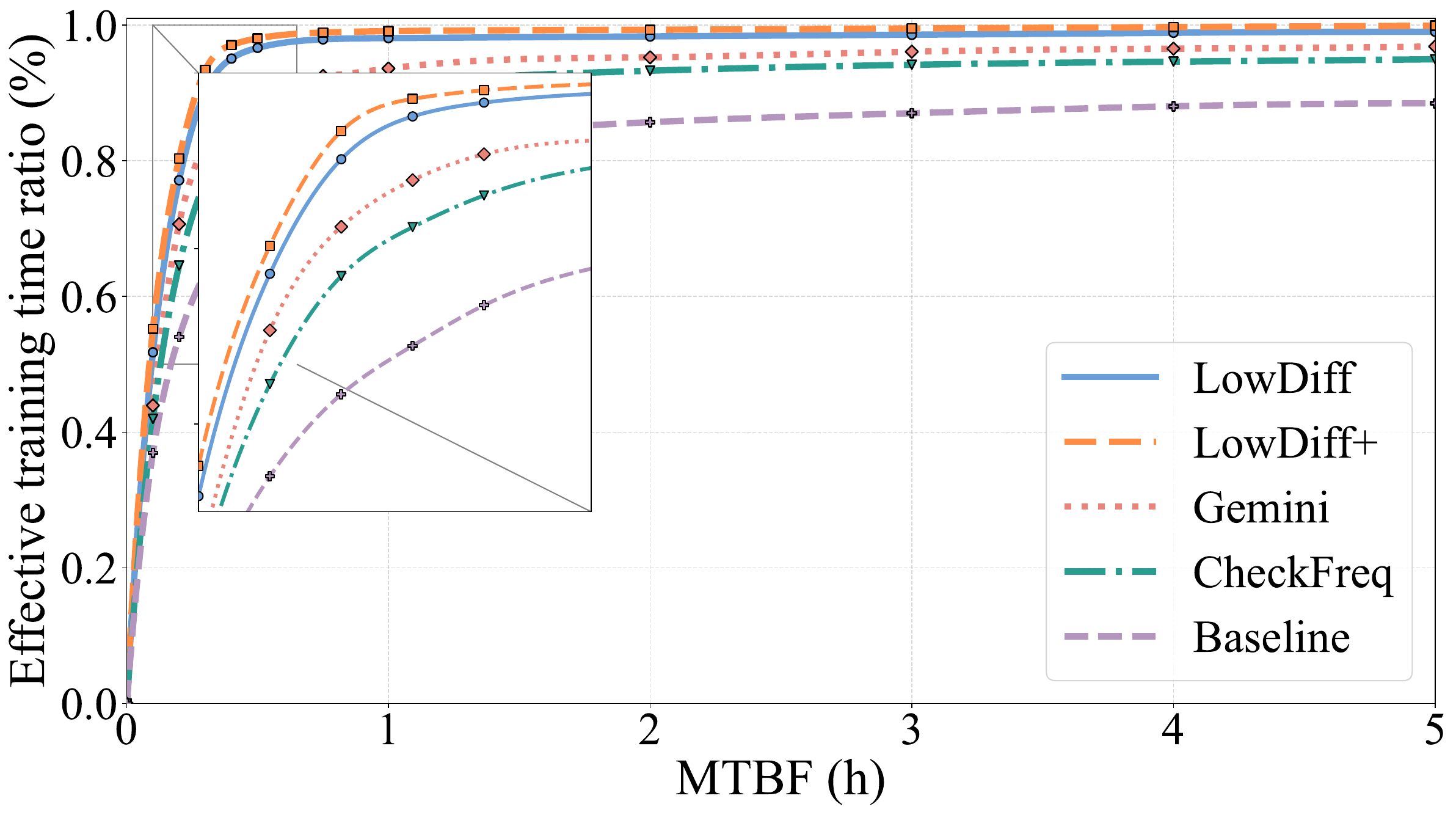}
        \end{minipage}
        &
        % ==================== Right: Exp. 11 ====================
        \begin{minipage}[t][1.55in][c]{0.48\textwidth}
            \vspace{0pt}
            \centering
            \includegraphics[width=\linewidth]{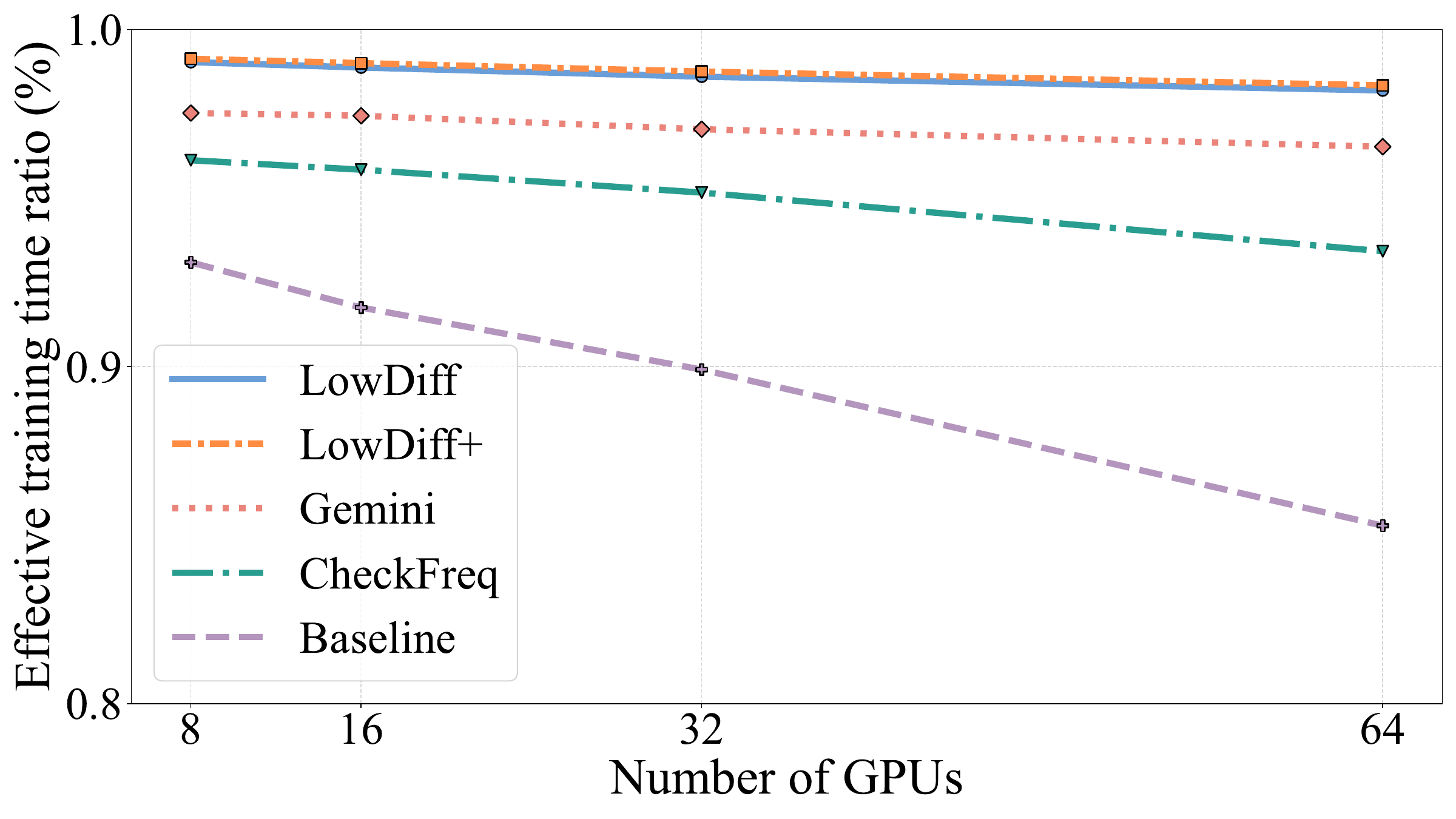}
        \end{minipage}

        \\[-6pt]

        % ==================== Captions ====================
        \begin{minipage}[t][0.38in][t]{0.48\textwidth}
            \vspace{0pt}
            \captionsetup{skip=0pt}
            \caption{(Exp.~10) Training under frequent software failures.}
            \Description{Training under frequent software failures.}
            \label{fig:impactfail}
        \end{minipage}
        &
        \begin{minipage}[t][0.38in][t]{0.48\textwidth}
            \vspace{0pt}
            \captionsetup{skip=0pt}
            \caption{(Exp.~11) Training with more GPUs.}
            \Description{More GPUs.}
            \label{fig:instance}
        \end{minipage}
    \end{tabular}
    \vspace{-20pt}
\end{figure*}

Figure~\ref{fig:impactfail} demonstrates that \sysnameplus consistently achieves the highest effective training time ratio across all failure scenarios, especially frequent ones. The effective training time ratio is determined by both steady checkpointing overhead and failure-induced recovery and recomputation costs. By recovering in memory, \sysnameplus reduces failure recovery overhead, while \sysname achieves the second-highest ratio by reducing checkpointing overhead through low-cost differential checkpointing. At an MTBF of 0.3 hours, \sysnameplus achieves an effective training time ratio of 93.3\%, followed by \sysname with 92.1\%.

\noindent\textbf{Scaling to more GPUs (Exp.~11).}  We evaluate the training performance by varying the number of GPUs used in the training process, using the same effective training time ratio metric as in Experiment 10. As the number of GPUs increases, the probability of failures in the cluster also rises \cite{gupta2024just}. We conduct experiments with 8, 16, 32, and 64 GPUs, respectively, and measure the effective training time ratio of \texttt{torch.save} (baseline), CheckFreq, Gemini, \sysname, and \sysnameplus. 

Figure \ref{fig:instance} shows that, as the number of GPUs increases, the effective training time ratio decreases; however, \sysname and  \sysnameplus still maintain a nearly 98\% proportion of the effective training time ratio, while the metric for other methods declines rapidly and can only reach around 90\%. This highlights the strong potential of \sysname and \sysnameplus in large-scale training.

\section{Conclusion}
In this paper, we identify gradient reuse as our main idea for low-cost differential checkpointing: \sysname reuses compressed gradients as differential checkpoints, eliminating redundant differential computation and reducing transmission cost, and further improves efficiency through batched gradient writing; \sysnameplus further extends gradient reuse to training systems without gradient compression via layer-wise-reuse snapshotting and incremental-merging persistence. Experiments on billion-parameter-scale models demonstrate that \sysname and \sysnameplus substantially reduce training overhead, wasted GPU time, and recovery latency under high-frequency checkpointing.
\bibliographystyle{ACM-Reference-Format}
\bibliography{refs}
\end{document}